\documentclass[aps,preprintnumbers,showpacs,nofootinbib,superscriptaddress,floatfix,notitlepage]{revtex4-1}

\pdfoutput=1 

\usepackage[english]{babel}
\usepackage{graphicx}
\usepackage{graphics}
\usepackage{braket}
\usepackage{bbold}
\usepackage{amssymb}
\usepackage{amsmath}
\usepackage{nicefrac}
\usepackage{dcolumn}
\usepackage{bm}
\usepackage{slashed}
\usepackage{datetime}
\usepackage{mciteplus}
\usepackage{multirow}
\usepackage{bbold}
\usepackage{siunitx}
\usepackage{booktabs}
\usepackage{color}
\usepackage{xcolor}
\definecolor{dgreen}{rgb}{0,0.325,0}
\definecolor{urlblue}{rgb}{0.2,0.4,0.7}
\definecolor{citegreen}{rgb}{0,0.4,0.2}
\definecolor{linkred}{rgb}{0.9,0.2,0.1}
\usepackage{float}
\usepackage[utf8]{inputenc}
\usepackage[normalem]{ulem}
\usepackage{fancyhdr}
\usepackage{hyperref}
\hypersetup{
 linktocpage = true,
 urlcolor = urlblue,
 colorlinks = true,
 linkcolor = linkred,
 anchorcolor = linkred,
 citecolor = linkred,
 pdfstartview = {XYZ null null 1.25} 
           }

\usepackage[normalem]{ulem} 

\newcommand{\tarr}{
\begin{array}}
\newcommand{\earr}{\end{array}}

\newcommand{\pt}{p_T}
\newcommand{\St}{S_T}
\newcommand{\pT}{\bm{p}_T}
\newcommand{\lT}{\bm{l}_T}
\newcommand{\kT}{\bm{k}_T}
\newcommand{\ST}{\bm{S}_T}
\newcommand{\epS}{\epsilon_T^{p_T S_T}}

\newcommand{\elS}{\epsilon_T^{l_T S_T}}
\newcommand{\LXtwoL}{L_X^2(\Lambda_X^2)}
\newcommand{\LXfourL}{L_X^4(\Lambda_X^2)}
\newcommand{\LXsixL}{L_X^6(\Lambda_X^2)}

\newcommand{\tr}{\operatorname*{Tr}\nolimits}

\begin{document}
\allowdisplaybreaks[2]

\title{T-odd gluon distribution functions in a spectator model}

\author{Alessandro Bacchetta}
\email{alessandro.bacchetta@unipv.it}
\affiliation{Dipartimento di Fisica, Universit\`a di Pavia, via Bassi
  6, I-27100 Pavia}
\affiliation{INFN Sezione di Pavia, via Bassi 6,
  I-27100 Pavia, Italy}

\author{Francesco Giovanni Celiberto}
\email{francesco.celiberto@uah.es}
\affiliation{Universidad de Alcal\'a (UAH), Departamento de F\'isica y Matem\'aticas, E-28805 Alcal\'a de Henares, Madrid, Spain}

\author{Marco Radici}
\email{marco.radici@pv.infn.it}
\affiliation{INFN Sezione di Pavia, via Bassi 6, I-27100 Pavia, Italy}

\begin{abstract}
We present a model calculation of T-odd transverse-momentum-dependent distributions of gluons in the nucleon. The model is based on the assumption that a nucleon can emit a gluon, 
and what remains after the emission is treated as a single spectator particle. This spectator particle is considered to be on-shell, but its mass is allowed to take a continuous range of values, described by a spectral function. The final-state interaction that is necessary to generate T-odd functions is modeled as the exchange of a single gluon between the spectator and the outgoing parton.
\end{abstract}

\date{\today, \currenttime}


\maketitle

\begingroup
 \hypersetup{linktoc = page, 
             }
 \vspace{-0.50cm}
 { 
 \tableofcontents
 }
 \phantom{.}\\\phantom{.}\\\phantom{.}
\endgroup

\setlength{\parskip}{3pt}%

\section{Introduction}
\label{s:intro}

The multi-dimensional distribution of partons within a nucleon can be parametrized in terms of several sets of functions, encoding different correlations between the momentum and spin of the parton and its parent nucleon. In particular, the 3-dimensional distribution in momentum space is encoded in the so-called Transverse-Momentum-Dependent Parton Distributions (TMD PDFs or TMDs)~\cite{Angeles-Martinez:2015sea}.
In simple terms, TMDs extend the conventional 1-dimensional collinear Parton Distribution Functions (PDFs) into three dimensions, including also the dependence on the partonic transverse momentum.

The endeavor to constrain TMDs is a crucial step toward unraveling the multi-dimensional structure of the nucleon, and gaining deeper insight into Quantum ChromoDynamics (QCD) and color confinement.  
The field of TMDs has witnessed remarkable advancement in recent years, predominantly in the quark sector. Progress within the gluon sector has been relatively restrained, owing to the challenges associated with probing gluons in high-energy processes.

Gluon TMDs at leading twist, first analyzed and classified in Ref.~\cite{Mulders:2000sh}, are shown in Tab.~\ref{tab:gluon_TMDs} in terms of both the polarization of the gluon and of its parent hadron. 
In this paper, our focus centers on (na\"ive) time-reversal odd (T-odd) gluon TMDs, highlighted in red in Tab.~\ref{tab:gluon_TMDs}. 
A notable example of a T-odd TMD is the gluon Sivers function, denoted as $f_{1T}^{\perp g}$. This function describes the distribution of unpolarized gluons in a transversely polarized nucleon and has a crucial role in the description of transverse single-spin asymmetries (see \cite{Boer:2015vso} and references therein).
As in the case for quark TMDs, T-odd gluon TMDs are generated by the presence of initial and/or final state QCD interactions between incoming or outgoing partons and the target fragments. These interactions also underlie the peculiar process-dependence of gluon TMDs.

{
\renewcommand{\arraystretch}{1.7}

\begin{table}[h]
 \centering
 \hspace{0.70cm} gluon polarization \\ \vspace{0.1cm}
 \rotatebox{90}{\hspace{-1.45cm} nucleon polarization} \hspace{0.1cm}
 \large
 \begin{tabular}[c]{|c|c|c|c|}
 \hline
 & $U$ & \; circular \; & linear \\
 \hline
 \; $U$ \; & $f_{1}^{g}$ & & \textcolor{blue}{$h_{1}^{\perp g}$} \\
 \hline	
 $L$ & & $g_{1}^{g}$ & \textcolor{red}{$h_{1L}^{\perp g}$} \\
 \hline	
 $T$ & \textcolor{red}{\; $f_{1T}^{\perp g}$ \;} & \textcolor{blue}{$g_{1T}^{g}$} & \; \textcolor{red}{$h_{1}^{g}$}, \textcolor{red}{$h_{1T}^{\perp g}$} \; \\
 \hline
  \end{tabular}
 \caption{Gluon TMD PDFs at twist-2. We adopt here the notation suggested in Ref.~\cite{Meissner:2007rx}, similar to the quark case. 
 $U$, $L$, $T$ depict unpolarized, longitudinally polarized and transversely polarized nucleons. 
 $U$, `circular', `linear' describe unpolarized, circularly polarized and linearly polarized gluons. 
 Functions in blue are T-even. 
 Functions in black are T-even and survive transverse-momentum integration. 
 Functions in red are T-odd. }
 \label{tab:gluon_TMDs}
\end{table}

}

Experimental information on gluon TMDs is very scarce, and particularly so for T-odd ones. Ref.~\cite{Lansberg:2017dzg} presented the first attempt to reconstruct the unpolarized gluon TMD, $f_1^g$. Phenomenological studies of the T-odd gluon Sivers function were published in Refs.~\cite{DAlesio:2017rzj,DAlesio:2018rnv,DAlesio:2020eqo}, but in processes where TMD factorization is not guaranteed to be applicable. An experimental measurement related to the gluon Sivers function was published by the COMPASS collaboration~\cite{COMPASS:2017ezz}. Several ways to experimentally access the gluon Sivers function have been discussed in the literature~\cite{Yuan:2008vn,Boer:2016fqd,Zheng:2018awe,Godbole:2014tha,Mukherjee:2016qxa,Bacchetta:2018ivt,DAlesio:2019qpk} and are among the primary goals of new experimental facilities~\cite{Boer:2011fh,Accardi:2012qut,Brodsky:2012vg,Aidala:2019pit}.

Pioneering calculations of gluon TMD distributions~\cite{Lu:2016vqu,Mulders:2000sh,Pereira-Resina-Rodrigues:2001eda} were performed using the \emph{spectator-model} approach (see also Refs.~\cite{Chakrabarti:2023djs,Xie:2022lra} for more recent versions). Originally conceived for studies in the quark-TMD sector~\cite{Bacchetta:2008af,Bacchetta:2010si,Gamberg:2005ip,Gamberg:2007wm,Jakob:1997wg,Meissner:2007rx}, this approach rests on the assumption that the struck nucleon emits a parton, and the residual fragments are treated as a single spectator particle, considered to be on-shell.  At variance with those studies, 
in Ref.~\cite{Bacchetta:2020vty} we presented the calculation of all T-even gluon TMDs in the spectator-model approach where 
the spectator mass is allowed to take a continuous range of values weighted by a flexible spectral function.
This modification encapsulates the effect of $q\bar{q}$ contributions, and 
allows 
to effectively reproduce both the small- and the moderate-$x$ behavior of the TMDs. 

In this paper, 
we extend the results of Ref.~\cite{Bacchetta:2020vty} by 
providing 
a systematic calculation 
in the same spectator-model framework of 
the complete set of 
all the 
four 
T-odd gluon TMDs at leading twist, 
including their process dependence. 

\section{The spectator model}
\label{s:Todd_gluon_correlator}

\begin{figure}[hb]
 \includegraphics[width=0.50\textwidth]{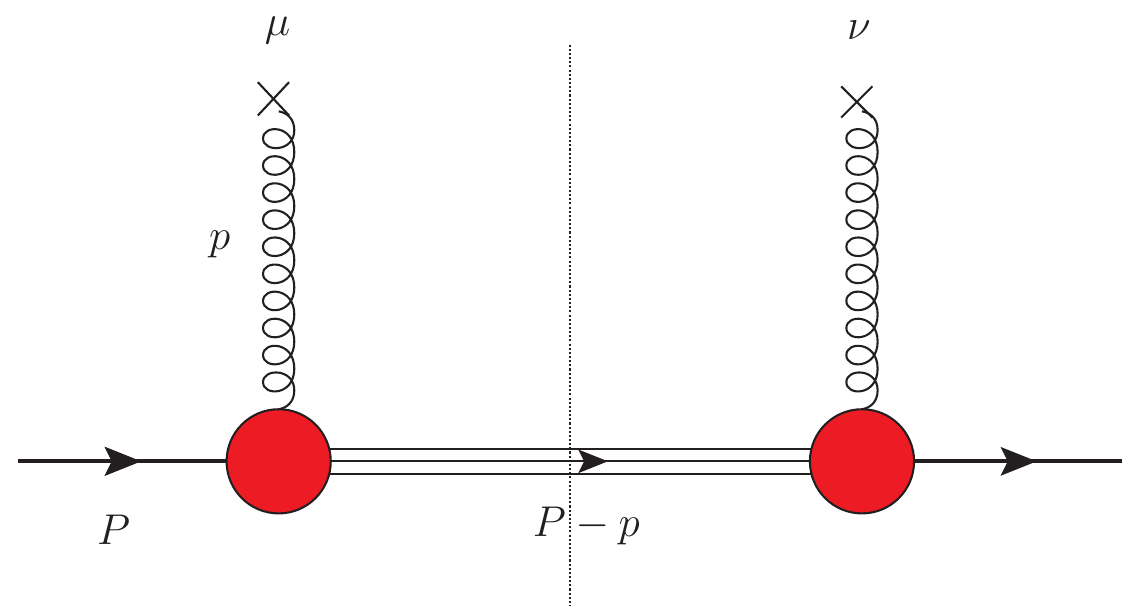}
 \caption{Tree-level cut diagram for the calculation of T-even leading-twist gluon TMDs. The triple line represents a spin-$\textstyle{\frac{1}{2}}$ spectator. The red blob represents the nucleon-gluon-spectator vertex.}
\label{fig:tree_level_diagram}
\end{figure}

Our model is 
based on the assumption that a nucleon can emit a gluon, and what remains after the emission is treated as a single spectator 
fermionic
particle (see Fig.~\ref{fig:tree_level_diagram}). This spectator 
fermion
is considered to be on-shell, but its mass is allowed to take a continuous range of values, described by a spectral function. The nucleon-gluon-spectator coupling is described by an effective vertex containing two form factors, 
inspired by the standard nucleon form factors. 
Such model can effectively reproduce the known collinear 
(un)polarized 
gluon PDFs
(the diagonal black entries $f_1^g$ and $g_1^g$ in Tab.~\ref{tab:gluon_TMDs}, that survive integration upon transverse momenta)
and can be used to compute all T-even TMDs~\cite{Bacchetta:2020vty}.

\begin{figure}[t]
 \centering
\includegraphics[width=0.65\textwidth]{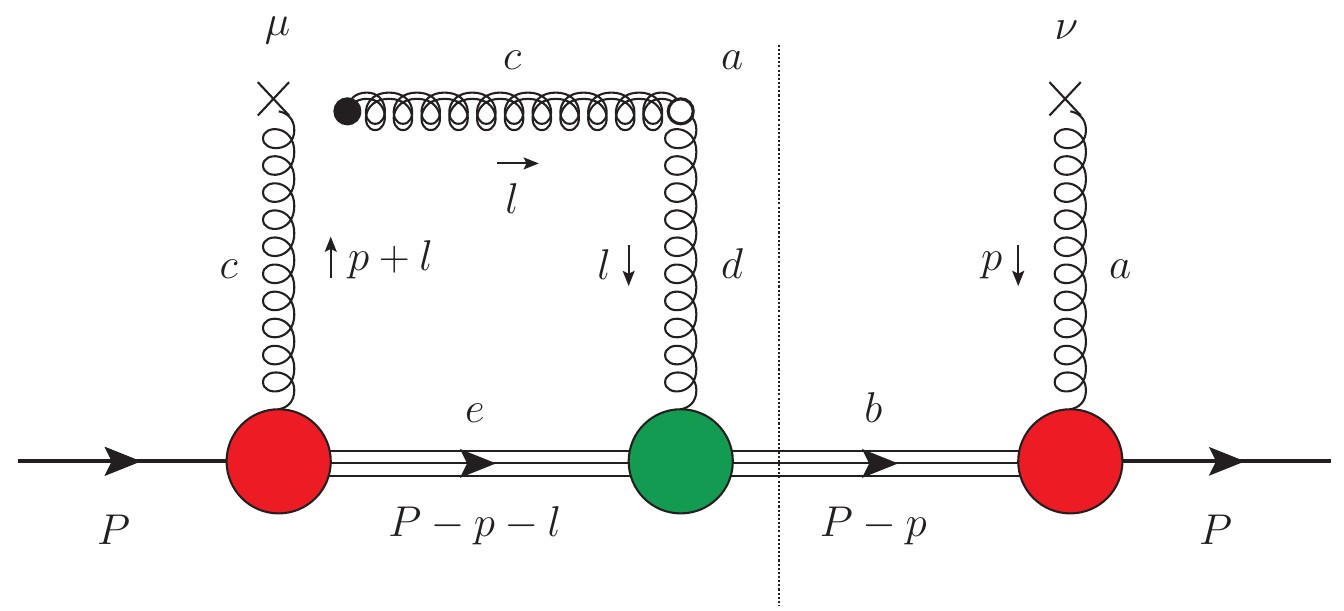} 
 \caption{Diagram for the calculation of the gluon-gluon correlator including the single-gluon exchange contribution, necessary to obtain T-odd TMDs.
 The eikonal propagator arising from the Wilson line in the operator definition of TMDs is indicated by a gluon double line. Only the imaginary part of the box diagram on the left-hand side of the cut is relevant for the calculation of T-odd functions. The red blobs represent the nucleon-gluon-spectator vertex with color indices $c e$ and $b a$, respectively, while the green blob stands for the spectator-gluon-spectator vertex with color indices $e d b$. The Hermitian-conjugate diagram is not shown.
 }
\label{fig:1-eik_Higgs_WW}
\end{figure}

T-odd gluon TMDs vanish at tree level, because there is no residual interaction between the active parton and the spectator; equivalently, there is no interference between two competing channels producing the complex amplitude whose imaginary part gives the T-odd contribution. We can generate such structures by considering the interference between the tree-level scattering amplitude and the scattering amplitudes with an additional gluon exchange, as shown in
Fig.~\ref{fig:1-eik_Higgs_WW}. This corresponds to the one-gluon-exchange approximation of the gauge link operator. As we shall discuss in detail, the exact form of the gauge link depends on the process and in our case leads to two different 
types of functions.

\subsection{Tree-level correlator}
\label{ss:correlator_LO}

Following Ref.~\cite{Bacchetta:2020vty}, we work in the frame where the nucleon momentum $P$ has no transverse component: 
\begin{equation}
P = \left[ \frac{M^2}{2 P^+}, \, P^+ , \, \bm{0} \right] \; ,
\label{eq:frame}
\end{equation}
where $M$ is the nucleon mass. The parton momentum is parameterized as 
\begin{equation}
p = \left[ \frac{p^2 + \pT^2}{2 x P^+} , \, x P^+ , \, \pT \right] \; ,
\label{eq:parton}
\end{equation}
where evidently $x = p^+ / P^+$ is the light-cone (longitudinal) momentum fraction carried by the parton.

In the spectator-model framework one assumes that the nucleon 
with spin $S$ 
in the state $|P, S \rangle$ can split into a gluon with momentum $p$ and other remainders, effectively treated as a single spin-$\textstyle{\frac{1}{2}}$ spectator particle with momentum $P-p$ and mass $M_X$. Similarly to Refs.~\cite{Bacchetta:2008af,Bacchetta:2020vty}, we define a ``tree-level"
correlator as (see Fig.~\ref{fig:tree_level_diagram})~\footnote{We remark that in Ref.~\cite{Bacchetta:2020vty} there is an error in the position of the ${\cal Y}$ vertices and a typo in the definition of the $G^{\mu \rho}$ propagator.}
\begin{equation}
\label{e:Phi}
\begin{split}
 \hspace{-0.15cm}
 \Phi^{\mu\nu (0)} (x, \pT, S) &=
 \frac{1}{(2 \pi)^3 \, 2 \, (1-x)\, P^+}
 \tr \Biggl[ (\slashed{P} + M) \,  \frac{1 + \gamma^5 \slashed{S}}{2} 
 \, G^{* \nu \sigma}(p,p) 
 \mathcal{Y}^{\dagger ab}_\sigma \big(p^2\big)
 (\slashed{P} - \slashed{p} + M_X)\, 
 \mathcal{Y}^{ba}_{\rho}(p^2)\, G^{\mu \rho}(p,p) 
  \Biggr],
\end{split}
\end{equation}
where $a$, $b$ are color indices (in the adjoint representation)
and
%
\begin{equation} 
  G^{\mu \rho} (p,q)
  = - \frac{i}{q^2 - m_g^2}\biggl(g^{\mu \rho} -\frac{ q^{\mu} (n_-)^{\rho}}{p^+}\biggr)
\end{equation}
is a specific Feynman rule for the gluon propagator in the definition of the correlator~\cite{Goeke:2006ef,Collins:2011zzd}, 
with $n^\rho_-$ a light-like unit vector of the light-cone basis, and $m_g$ a gluon mass regulator which will be set to zero in our calculations. 
We model the nucleon-gluon-spectator vertex as
\begin{equation}
 \label{eq:vertex_ngs}
 \mathcal{Y}^{ba}_\rho(p^2) = \delta^{ba} \, \bigg[ g_1 (p^2) \, \gamma_\rho + g_2 (p^2) \, \frac{i}{2 M}\, \sigma_{\rho \nu} \, p^\nu \bigg] \, , 
\end{equation}
where as usual $\sigma_{\rho \nu} = i [ \gamma_\rho , \gamma_\nu ] /2$, and $g_{1,2} (p^2)$ are generic form factors. In principle, the expression
of $\mathcal{Y}^{ba}_\rho(p^2)$ could contain more Dirac structures. However, with our assumptions the spectator is identified with an on-shell
spin-$\textstyle{\frac{1}{2}}$ particle, much like the nucleon. Hence, we model the structure of $\mathcal{Y}^{ba}_\rho(p^2)$ similarly to the conserved
electromagnetic current of a free nucleon obtained from the Gordon decomposition. The form factors $g_{1,2} (p^2)$ are formally similar to the
Dirac and Pauli form factors, but obviously must not be identified with them. Similarly to our previous model description of quark TMDs~\cite{Bacchetta:2008af}, we use the dipolar expression 
\begin{equation}
g_{1,2} (p^2) = \kappa_{1,2} \, \frac{p^2}{|p^2 - \Lambda_X^2|^2} = \kappa_{1,2} \, \frac{p^2 \, (1 - x)^2}{(\pT^2 + L_X^2 (\Lambda_X^2))^2}  \ ,
\label{eq:dipolarff}
\end{equation}
where $\kappa_{1,2}$ and $\Lambda_X$ are normalization and cut-off parameters, respectively, and 
\begin{equation}
L_X^2 (\Lambda_X^2) = x \, M_X^2 + (1 - x) \, \Lambda_X^2 - x \, (1 - x) \, M^2 \, .
\label{eq:LX}
\end{equation}
The dipolar expression of Eq.~\eqref{eq:dipolarff} has several advantages: it cancels the singularity of the gluon propagator, it smoothly suppresses the effect of high $\pT^2$ where the TMD formalism cannot be applied, and it compensates also the logarithmic divergences arising after integration upon $\pT$.

In our model, 
the overall 
color 
prefactor at tree level 
is
\begin{equation}
    C^{(0)} = \delta^{ab} \delta^{ba} = 8. 
\end{equation}

As a comparison, we will also discuss the quark-target model, which can be obtained from Eq.~\eqref{e:Phi} simply by replacing
\begin{align}
\mathcal{Y}_{\rho}^{b a}  & \to 
g_s \gamma_{\rho} t^{a} \; , 
\end{align}
with $g_s$ the strong coupling constant and $t^a$ a generator of color SU(3) transformations, 
and by setting $M = M_X \equiv m_q$ everywhere.
In this case, the overall color factor is
\begin{equation}
    C_q^{(0)} = \frac{1}{N_C}\tr_C[ t^a t^a ]= \frac{4}{3}, 
\end{equation}
where $N_C$ is the number of colors and $\tr_C$ indicates the trace upon color indices.


\subsection{Additional single-gluon exchange}
\label{ss:process_dependence_gluon_TMDs}


In general, T-odd TMDs arise only when there is a  residual interaction between the active parton and the spectator. More specifically, they arise from the imaginary part of the interference between the tree-level channel and the channel describing this residual interaction. Following our model calculation for quark TMDs~\cite{Bacchetta:2008af}, we generate this interference by describing the residual gluon-spectator interaction through the exchange of a soft gluon (see Fig.~\ref{fig:1-eik_Higgs_WW}). This one-gluon exchange results from the truncation at the first order in the expansion of the path-ordered exponential that defines the gauge link as the sum of infinite gluon rescatterings~\cite{Efremov:1978xm}. 

In the general definition of the parton-parton correlator $\Phi$, the gauge link is a necessary ingredient to make the correlator color-gauge invariant. 
However, the sensitivity of TMDs to the transverse components of the gauge link introduces a process dependence, contrary to the case of collinear PDFs. While T-even quark TMDs are independent from the direction of the color flow in the involved hard scattering, T-odd quark TMDs change sign when moving from final-state interactions with future-pointing $([+])$ Wilson lines (like in Semi-Inclusive Deep-Inelastic Scattering - SIDIS) to initial-state interactions with past-pointing $([-])$ Wilson lines (like in Drell-Yan processes)~\cite{Brodsky:2002cx,Collins:2002kn,Belitsky:2002sm}. 

The gluon case is more intricate, due its color-octet structure, and leads 
to a more diversified form of modified universality with respect to the quark case. 
There is a gauge link with color flowing through a closed path pointing to the future, corresponding to final-state interactions between the spectator and an outgoing gluon, like in SIDIS production of two jets or heavy-quark pairs~\cite{Pisano:2013cya}. This gauge link 
is usually denoted with the $[+,+]$ symbol. Conversely, initial-state interactions are described by gauge links with past-pointing close Wilson lines $([-,-])$ and occur, for example, in Higgs production via gluon fusion ($gg \to H$)~\cite{Boer:2013fca,Echevarria:2015uaa}. The gluon TMDs originating from these gauge links are called 
Weizs\"acker--Williams (WW) gluon TMDs, or $f$-type gluon TMDs because the color structure of the T-odd ones involves the antisymmmetric structure constants $f$ of the color gauge group SU(3). It turns out that T-even WW gluon TMDs are symmetric with respect to the different paths $([+,+] = [-,-])$, while the T-odd WW ones change sign.  

Moreover, color can flow through a closed path involving both initial and final states, like in photon-jet production from hadronic collisions or SIDIS~\cite{Bomhof:2006dp,Bacchetta:2007sz,Buffing:2018ggv}. We remark that for this class of processes TMD factorization is not expected to hold~\cite{Rogers:2013zha}; however, it is still possible to calculate the corresponding TMDs in the context of our model. Depending on the direction of color flow, we have $[+,-]$ and $[-,+]$ structures and the corresponding gluon TMDs are usually called dipole gluon TMDs, or $d$-type gluon TMDs because their T-odd color structure involves the symmetric structure constants $d$ of color SU(3).~\footnote{Due to the connection between the T-odd TMDs at twist-2 and the collinear PDFs at twist-3, the distinction between $f$-type and $d$-type gluon TMDs appears already in the correlator of the Qiu--Sterman twist-3 collinear PDF~\cite{Qiu:1991pp,Qiu:1991wg,Qiu:1998ia}.} Similarly to the WW case, the dipole T-even gluon TMDs are symmetric with respect to different color paths $([+,-] = [-,+])$, while T-odd dipole ones change sign. But, more importantly, WW and dipole gluon TMDs are not related to each other, and contain different physical information.

We first compute the gluon-gluon correlator corresponding to the $[+,+]$ gauge link with future-pointing closed Wilson path. The one-gluon exchange approximation of the gauge link amounts to compute the diagram depicted in Fig.~\ref{fig:1-eik_Higgs_WW}. The double gluon line represents the struck gluon described in the eikonal approximation, following the same procedure of the quark case~\cite{Bacchetta:2008af}. 
The Feynman rules to describe the eikonal gluon line and the eikonal vertex are written in detail in Ref.~\cite{Buffing:2017mqm}.

The expression of the correlator turns out to be 
\begin{equation}
\label{e:Phi++}
\begin{split}
 \Phi^{\mu\nu [+,+]} &(x, \pT, S) =
 \frac{1}{(2 \pi)^3 \, 2 \, (1-x)\, P^+}
 \tr \Biggl[ 
 (\slashed{P} + M) \,  \frac{1 + \gamma^5 \slashed{S}}{2} 
 \, G^{* \nu \sigma}(p,p) 
 \mathcal{Y}^{\dagger ab}_\sigma \big(p^2\big)
 (\slashed{P} - \slashed{p} + M_X)\, 
 (g_s {n}_-^{\alpha} f^{dac})
 \\
& \times \, \int \frac{d^4 l}{(2\pi)^4} \biggl( \frac{-i
 \mathcal{X}^{bde}_{\alpha}(l^2)}{l^2 - m_g^2} \biggr) \biggl( \frac{- i }{l^+ + i \epsilon}\biggr) \, \frac{i \big(\slashed{P} - \slashed{p} - \slashed{l} + M_X\big)}{(P - p - l)^2 - M_X^2 + i \epsilon} \,
 \mathcal{Y}^{ec}_{\rho}\big((p + l)^2\big)
  \, G^{\mu \rho}(p,p+l) 
 \Biggr] \;,
\end{split}
\end{equation}
where $\mathcal{X}^{bde}_\alpha$ 
is the spectator-gluon-spectator vertex 
to be defined in Section~\ref{ss:vertex_sns}.

The correlator $\Phi^{\mu \nu [-,-]}$ for the $[-,-]$ past-pointing closed Wilson path can be obtained by changing 
the sign of the $+i \epsilon$ term in Eq.~\eqref{e:Phi++}.


The correlator $\Phi^{\mu \nu [+,-]}$ for the $[+,-]$ gauge link (leading to $d-$type gluon TMDs~\cite{Kharzeev:2003wz,Dominguez:2010xd,Dominguez:2011wm}) can be simply derived by replacing in the eikonal vertex the antisymmetric color structure $f^{dac}$ with the symmetric $-i d^{dac}$ in Eq.~\eqref{e:Phi++}:
\begin{equation}
\label{e:Phi+-}
\begin{split}
 \Phi^{\mu\nu [+,-]} &(x, \pT, S) =
 \frac{1}{(2 \pi)^3 \, 2 \, (1-x)\, P^+}
 \tr \Biggl[ 
 (\slashed{P} + M) \,  \frac{1 + \gamma^5 \slashed{S}}{2} 
 \, G^{* \nu \sigma}(p,p) 
 \mathcal{Y}^{\dagger ab}_\sigma \big(p^2\big)
 (\slashed{P} - \slashed{p} + M_X)\, 
 (-i g_s {n}_-^{\alpha} d^{dac})
 \\
& \times \, \int \frac{d^4 l}{(2\pi)^4} \biggl( \frac{-i
 \mathcal{X}^{bde}_{\alpha}(l^2)}{l^2 - m_g^2} \biggr) \biggl( \frac{- i }{l^+ + i \epsilon}\biggr) \, \frac{i \big(\slashed{P} - \slashed{p} - \slashed{l} + M_X\big)}{(P - p - l)^2 - M_X^2 + i \epsilon} \,
 \mathcal{Y}^{ec}_{\rho}\big((p + l)^2\big)
  \, G^{\mu \rho}(p,p+l) 
 \Biggr] \;.
\end{split}
\end{equation}

As 
for the WW 
case, 
the $[-,+]$ correlator 
differs from the $[+,-]$ one only by 
the sign of the $+i \epsilon$ term in Eq.~\eqref{e:Phi+-}.

Our model agrees with the relations between gluon TMDs with different gauge link structures that have been systematically studied in~\cite{Buffing:2013kca}. For example, for the T-even unpolarized function, $f_1^g$, and for the T-odd gluon Sivers function, $f_{1T}^{g\,\perp}$, one has the following modified-universality relations~\cite{Bomhof:2006dp,Buffing:2013kca,Boer:2016fqd}:
\begin{alignat}{2} 
    f_1^{g\,[+,+]} &= f_1^{g\,[-,-]} \;, \qquad& f_1^{g\,[+,-]} &= f_1^{g\,[-,+]} \;, \\
    f_{1T}^{\perp \, g \, [+,+]} &= -f_{1T}^{\perp \, g \, [-,-]} \;, \qquad& f_{1T}^{\perp \, g \, [+,-]} &= -f_{1T}^{\perp \, g \, [-,+]} \;. \label{eq:gluon_Sivers}
\end{alignat}
As it turns out, in general $f_1^{g\,[+,+]}$ \emph{cannot} be related to $f_1^{g\,[+,-]}$, and likewise for $f_{1T}^{\perp\, g}$. 
They 
encode 
different
information and require different extractions~\cite{Boer:2015vso}.

\subsection{Spectator-gluon-spectator vertex}
\label{ss:vertex_sns}

A key ingredient 
of our model
is the spectator-gluon-spectator vertex $\mathcal{X}^{bde}_\alpha$, depicted by a green 
blob in Fig.~\ref{fig:1-eik_Higgs_WW}. 
If the nucleon-gluon-spectator vertex $\mathcal{Y}^{ec}_\mu$ (red blob) 
connects 
a colorless initial-state (nucleon) to an octet state (gluon) and an anti-octet state (spectator), the spectator-gluon-spectator vertex $\mathcal{X}^{bde}_\alpha$ connects an anti-octet initial state (spectator) to an octet state (gluon) and an anti-octet state (spectator). Since 
in our model 
the spectator 
is assumed to be a spin-$\textstyle{\frac{1}{2}}$ particle describing 
a collection of partons 
as remainders, 
the 
vertex $\mathcal{X}^{bde}_\alpha$ can in principle contain both the $f^{bde}$ and $d^{bde}$ color structure constants, 
each one multiplying a Dirac structure similar to Eq.~\eqref{eq:vertex_ngs}:
\begin{equation}
 \label{eq:vertex_sgs}
 \mathcal{X}^{bde}_\alpha(p^2) = f^{bde} \, \bigg[ g_1^f (p^2) \, \gamma_\alpha + g_2^f (p^2) \, \frac{i}{2 M}\, \sigma_{\alpha \beta} \, p^\beta \bigg] -i \, d^{bde} \, \bigg[ g_1^d (p^2) \, \gamma_\alpha + g_2^d (p^2) \, \frac{i}{2 M}\, \sigma_{\alpha \beta} \, p^\beta \bigg] \; , 
\end{equation}
where $g^{f,d}_{1,2} (p^2)$ are \emph{a priori} four different functions of $p^2$.
In principle, they are independent from the $g_{1,2} (p^2)$ form factors entering the 
nucleon-gluon-spectator vertex of Eq.~\eqref{eq:vertex_ngs}. 
For the sake of simplicity, 
we will assume 
$g^{d}_{1,2} (p^2) = g^{f}_{1,2} (p^2)$ 
and we will get 
$d$-type densities 
equal to the corresponding $f$-type ones up to a color factor:
\begin{align}
C^{[+,+]}    &= f^{acd} f^{dca} = - 2 C_A^2 C_F = - 24 \; , \\
C^{[+,-]}    &= (- i\, d^{acd})\, (-i\, d^{dca}) = 2 \left(4-C_A^2\right) C_F = - \frac{40}{3} \; .
\label{e:color_factor}
\end{align}
For this reason, in the following we will show results only for $f$-type gluon TMDs, and we will drop the $[+,+]$ index when not needed. 

In the quark-target model, we would 
replace the spectator-gluon-spectator vertex by $\mathcal{X}^{bde}_\alpha \to (-i g_s \gamma_{\alpha} t^{d} )$ and obtain the color factors 
\begin{align}
C^{[+,+]}_q    &= -\frac{i}{N_C}\tr_C[ t^a t^c t^d] f^{acd}=
\frac{C_A C_F}{2} = -2 \; , \\
C^{[+,-]}_q    &= -\frac{1}{N_C}\tr_C[ t^a t^c t^d] d^{acd}=
\frac{C_F}{2} \left(4-C_A^2\right) \left(C_A-2 C_F\right)=
-\frac{10}{9} \; .
\end{align}
Note that the ratio of the two different gauge link structures remains the same in the two cases: $[+,-] / [+,+] = 5/9$. With our simplified assumptions, therefore, the T-odd $d$-type functions are always about half of the $f$-type ones.

We further assume 
\begin{equation}
 \label{eq:g}
 g^{d}_{1,2} (p^2) = g^{f}_{1,2} (p^2) \equiv g_{1,2} (p^2) \;. 
\end{equation}
This means that the parameters entering our model for $f$-type and $d$-type T-odd gluon TMDs are fully determined by those ones entering the T-even gluon TMDs that contain $g_{1,2} (p^2)$ through Eq.~\eqref{eq:vertex_ngs}. The latter parameters have been fixed by fitting the 
integrated T-even gluon TMDs on the known corresponding collinear PDFs~\cite{Bacchetta:2020vty} (see Tab.~\ref{tab:par-g}).

\subsection{Gluon TMD projectors}
\label{ss:projectors}

T-odd gluon TMDs 
can be extracted from the analytic structure of the gluon-gluon correlator by making use of suitable projectors. Using Eqs.(52-54) of Ref.~\cite{Meissner:2007rx} for the general parametrization of the gluon-gluon correlator $\Phi^{\mu\nu} (x, \pT, S)$ for three different nucleon polarizations $S=0, S_L, \ST$, it is possible to show that the four T-odd gluon TMDs of Tab.~\ref{tab:gluon_TMDs} can be isolated through the following projections: 
\begin{align}
 f_{1T}^{\perp g} &= \mathbb{P}_{[f_{1T}^{\perp g}]}^{\mu \nu} \, \left[ \Phi_{\mu\nu} (x, \pT, \ST) - \Phi_{\mu\nu} (x, \pT, -\ST) \right] \nonumber \\ 
 &= \frac{M}{2} \frac{1}{\epS} g_T^{\mu\nu} \, \left[ \Phi_{\mu\nu} (x, \pT, \ST) - \Phi_{\mu\nu} (x, \pT, -\ST) \right]\;, \label{eq:projectors_SIV} \\[0.2cm]
 h_{1T}^{\perp g} &= \mathbb{P}_{[h_{1T}^{\perp g}]}^{\mu \nu} \, \left[ \Phi_{\mu\nu} (x, \pT, \ST) - \Phi_{\mu\nu} (x, \pT, -\ST) \right] \nonumber \\ 
 &=  \frac{M^3}{\pt^2} \frac{1}{\epS} \left( \frac{p_T^{\{\mu} S_T^{\nu\}}}{\pt\cdot\St} - g_T^{\mu\nu} \right) \, \left[ \Phi_{\mu\nu} (x, \pT, \ST) - \Phi_{\mu\nu} (x, \pT, -\ST) \right] \;, \label{eq:projectors_PRETZ} \\[0.2cm]
 h_1^g &= \mathbb{P}_{[h_1^g]}^{\mu \nu} \, \left[ \Phi_{\mu\nu} (x, \pT, \ST) - \Phi_{\mu\nu} (x, \pT, -\ST) \right] \nonumber \\ 
 &= \frac{M}{2 \epS} \, \left[ \frac{4}{\pt^2} \, \pt^\mu \pt^\nu + \frac{p_T^{\{\mu} S_T^{\nu\}}}{\pt\cdot\St} - 3 g_T^{\mu\nu} \right] \, \left[ \Phi_{\mu\nu} (x, \pT, \ST) - \Phi_{\mu\nu} (x, \pT, -\ST) \right] \;, \label{eq:projectors_LIN} \\[0.2cm]
 h_{1L}^{\perp g} &= \mathbb{P}_{[h_{1L}^{\perp g}]}^{\mu \nu} \, \left[ \Phi_{\mu\nu} (x, \pT, S_L) - \Phi_{\mu\nu} (x, \pT, - S_L) \right] \nonumber \\ 
 &= \frac{1}{S_L} \frac{M^2}{2 \pt^4} {\epsilon_T^{\{\mu}}_\alpha \pt^{\nu\}\alpha}\, \left[ \Phi_{\mu\nu} (x, \pT, S_L) - \Phi_{\mu\nu} (x, \pT, - S_L) \right] \;,
\label{eq:projectors_WG}
\end{align} 
where $\epsilon_T^{v w} \equiv \epsilon^{- + i j}\, v_i\, w_j$ with $i,j$ transverse spatial indices and $\epsilon^{\mu \nu \alpha \beta}$ the antisymmetric Levi-Civita tensor, and 
\begin{align}
\pt^{\mu\nu} &= \pt^\mu \pt^\nu - \frac{1}{2} \pt^2 g_T^{\mu\nu} \; , \\
g_T^{\mu \nu} &= g^{\mu \nu} - n_+^{\left\{ \mu \right.} n_-^{\left. \nu \right\}} \; , \\
v^{\left\{ \mu \right.} w^{\left. \nu \right\}} &= v^\mu w^\nu + v^\nu w^\mu \; .
\end{align}

\section{T-odd gluon TMDs: Illustrative examples}
\label{s:TMDs_g1}

\subsection{Sivers function: $g_1$-vertex approximation} 
\label{ss:Sivers_f_g1}


Let us consider first the $f$-type Sivers function $f_{1T}^{\perp \, [+,+]}$ with a simpler expression for the nucleon-gluon-spectator vertex, where the term proportional to $\sigma^{\mu\nu} p_\nu$ in Eq.~\eqref{eq:vertex_ngs} is neglected. In other words, the $g_2(p)$ coupling is set to zero 
and the vertex reduces to
\begin{equation}
\mathcal{Y}^{ba}_{\rho} \to \delta^{b a} \, g_1 (p^2) \, \gamma_\rho \;.
 \label{eq:vertex_ngs_g1}
\end{equation}
We name this the ``$g_1$-vertex approximation". We indicate the resulting Sivers function as $f_{1T}^{\perp \, (g_1)}$, and similarly for all other TMDs computed in this approximation. 

Using for $g_1(p^2)$ the dipolar form of  Eq.~\eqref{eq:dipolarff}, the corresponding  projector of Eq.~\eqref{eq:projectors_SIV} specialized to a $f$-type gluon Sivers and applied to the correlator of Eq.~\eqref{e:Phi++}, we obtain
\begin{equation}
\begin{split}
 &
 f_{1T}^{\perp \, (g_1)}(x,\pT) = 
 \mathbb{P}_{[f_{1T}^{\perp g}]}^{\mu \nu} \, \left[ \Phi_{\mu\nu}^{[+,+]\, (g_1)} (x, \pT, \ST) - \Phi_{\mu\nu}^{[+,+] \, (g_1)} (x, \pT, -\ST) \right]
\label{f1Tp_pp_g1_1}
\\
 &=  \frac{48\,g_s \kappa_1^3}{(2\pi)^3} \frac{M \left[ M_X - M (1-x) \right]\,P^+}{(p^2 - \Lambda_X^2)^2} \; 2 \, {\rm Re} \int \frac{d^4 l}{(2\pi)^4} \frac{\elS}{(l^2 - \Lambda_X^2)^2} \frac{1}{l^+ + i\epsilon} \frac{1}{(l + p - P)^2 - M_X^2 + i\epsilon} \frac{1}{[(p + l)^2 - \Lambda_X^2 + i\epsilon]^2} \;. 
\end{split}
\end{equation}

In Eq.~\eqref{f1Tp_pp_g1_1}, 
terms proportional to $\epsilon^{l n_- p P} \equiv \epsilon^{\mu \nu \alpha \beta} l_\mu n_{-\nu} p_\alpha P_\beta$ and $\epsilon^{l p P S}$ vanish because 
the only component of $l$ contributing to the integral 
is the one parallel to $p$. 

Similarly to the calculation of the quark Sivers TMD~\cite{Bacchetta:2008af,Brodsky:2002rv,Gamberg:2003ey,Boer:2002ju,Bacchetta:2003rz}, the non vanishing contribution to the integral of Eq.~\eqref{f1Tp_pp_g1_1} comes from the poles of the two 
$[l^+ + i\epsilon]$ and $[(l + p - P)^2 - M_X^2 + i\epsilon]$ propagators. 

Using the Cutkosky's rules, we can make the replacement
\begin{equation}
 \frac{1}{l^+ + i\epsilon} \to -2 \pi i \, \delta(l^+) \;, \qquad \frac{1}{(l + p - P)^2 - M_X^2 + i\epsilon} \to -2 \pi i \, \delta((l + p - P)^2 - M_X^2) \;.
\label{prop_delta_f1Tp_pp}
\end{equation}
Moreover, we can 
also make use of the 
spectator model relation
\begin{equation}
 k^2 - \Lambda_X^2 = - \frac{\kT^2 + L_X^2(\Lambda_X)}{1-x} \;,
\label{p2_LX2}
\end{equation}
where $k^2$ generically refers to $p^2$, $l^2$ or $(p+l)^2$, and $\kT^2$ to the corresponding euclidean transverse parts.

The final result 
for the WW gluon Sivers function with only $g_1$ coupling is
\begin{equation}
 f_{1T}^{\perp \, (g_1)}(x,\pT) = - \frac{48\,g_s \kappa_1^3}{(2\pi)^3} \frac{M \left[ M_X - M (1-x) \right]\,(1-x)^5\,P^+}{[\pT^2 + L_X^2(\Lambda_X)]^2} \, {\cal D}_2 (p) \;,
\label{f1Tp_pp_g1_2}
\end{equation}
where
\begin{equation}
 {\cal D}_2 (p) = \frac{1}{2P^+} \int \frac{d^2 \lT}{(2\pi)^2} \frac{\lT \cdot \pT}{\pT^2}\frac{1}{[\lT^2 + \LXtwoL]^2} \frac{1}{[(\lT + \pT)^2 + \LXtwoL]^2} \;.
\label{D2_1}
\end{equation}

Introducing the Feynman parametrization, we can rewrite the integral as
\begin{equation}
 {\cal D}_2 (p) = \frac{1}{2 P^+} \int \frac{d^2 \lT}{(2\pi)^2} \frac{\lT \cdot \pT}{\pT^2} \int_0^1 d \alpha \, \frac{6\,\alpha\,(1-\alpha)}{\{\alpha\,[\lT^2 + \LXtwoL] + (1-\alpha)\,[(\lT + \pT)^2 + \LXtwoL]\}^4} \;.
\label{D2_2}
\end{equation}
After the change of variable $\lT \to \lT' = \lT + (1 - \alpha) \pT$, we have 
\begin{equation}
\begin{split}
 {\cal D}_2 (p) &= - \frac{1}{2P^+} \int_0^1 d \alpha \int \frac{d^2 \lT'}{(2\pi)^2} \, \frac{6 \, \alpha \, (1 - \alpha)^2}{[\lT'^2 + \alpha (1 - \alpha) \pT^2 + \LXtwoL]^4}
 \\ 
 &= - \frac{1}{4 \pi P^+} \int_0^1 d \alpha \, \frac{\alpha \, (1 - \alpha)^2}{[\alpha (1 - \alpha) \pT^2 + \LXtwoL]^3}
 \\ 
 &= - \frac{1}{8 \pi P^+} \left[ \frac{1 - 2 \, \LXtwoL/\pT^2}{\LXtwoL \, [\pT^2 + 4 \LXtwoL]^2} + 8 \frac{\pT^2 + \LXtwoL}{|\pT|^3 \, [\pT^2 + 4 \LXtwoL]^{\nicefrac{5}{2}}} \, \tanh^{-1} \left( \frac{|\pT|}{\sqrt{\pT^2 + 4 \LXtwoL}} \right) \right]\;.
\label{D2_3}
\end{split}
\end{equation}
Combining Eqs.~\eqref{f1Tp_pp_g1_2} and~\eqref{D2_3} we get the final expression for our $f$-type Sivers function in the $g_1$-vertex approximation
\begin{equation}
\begin{split}
 f_{1T}^{\perp \, (g1)}&(x,\pT) = \frac{12\,g_s \kappa_1^3}{(2\pi)^4} \frac{M \left[ M_X - M (1-x) \right]\,(1-x)^5}{[\pT^2 + L_X^2(\Lambda_X)]^2} \\
 & \quad \times \, \left[ \frac{1 - 2 \, \LXtwoL/\pT^2}{\LXtwoL \, [\pT^2 + 4 \LXtwoL]^2} + 8 \frac{\pT^2 + \LXtwoL}{|\pT|^3 \, [\pT^2 + 4 \LXtwoL]^{\nicefrac{5}{2}}} \, \tanh^{-1} \left( \frac{|\pT|}{\sqrt{\pT^2 + 4 \LXtwoL}} \right) \right] \;.
\label{f1Tp_pp_g1}
\end{split}
\end{equation} 



In order to explore the effects of the $g_1$-vertex approximation, we fix the model parameters by simultaneously fitting the integrated unpolarized and helicity gluon TMDs onto the corresponding known collinear PDFs. Following the methodology of Ref.~\cite{Bacchetta:2020vty}, we first allow the spectator mass to take a continuous range of values by weighting the gluon TMDs with the spectral function described in Eqs.(16,17) of Ref.~\cite{Bacchetta:2020vty}, which is a way to effectively take into account $q \bar{q}$ contributions. Then, we integrate the gluon TMDs upon the transverse momenta and we fix all the model parameters by fitting the unpolarized collinear PDF from {\tt NNPDF3.1sx}~\cite{Ball:2017otu} and the helicity collinear PDF from {\tt NNPDFpol1.1}~\cite{Nocera:2014gqa} at the indicated initial scale $Q_0 = 1.64$~GeV and in the range $10^{-3} < x < 0.7$.~\footnote{The $x > 0.7$ tail was excluded to avoid large uncertainties~\cite{NNPDF:2014otw} due to threshold effects and target-mass corrections, not accounted for in our model.} The only exception is the parameter $\kappa_2$ in Eq.~\eqref{eq:dipolarff} that controls the strength of the $g_2$ coupling; here, it is systematically set to zero. Statistical uncertainties are generated using the replica method,
widely used in the phenomenological extraction of quark densities from experimental data~\cite{Forte:2002fg,Ball:2017otu,Bacchetta:2017gcc,Scimemi:2019cmh,Bacchetta:2019sam,Bacchetta:2022awv,Bury:2022czx,Moos:2023yfa}.

In Tab.~\ref{tab:par-g}, we compare the obtained values (labelled {\tt PVGlue20g1V} in the two rightmost columns) with the original values from Ref.~\cite{Bacchetta:2020vty} (labelled {\tt PVGlue20} in the second and third columns from left). The 68\% uncertainties accompanying the central values are obtained by excluding the largest and smallest 16\% of all 100 replica values, which would correspond to $1 \sigma$ standard deviation for a Gaussian distribution. The columns labelled with ``replica 11" show the parameters of the most representative replica, because in both fits its parameter values have the minimal distance from the mean values.

\begin{figure}[t]
 
 \centering

 \hspace{0.090cm}
 \includegraphics[width=0.47\textwidth]{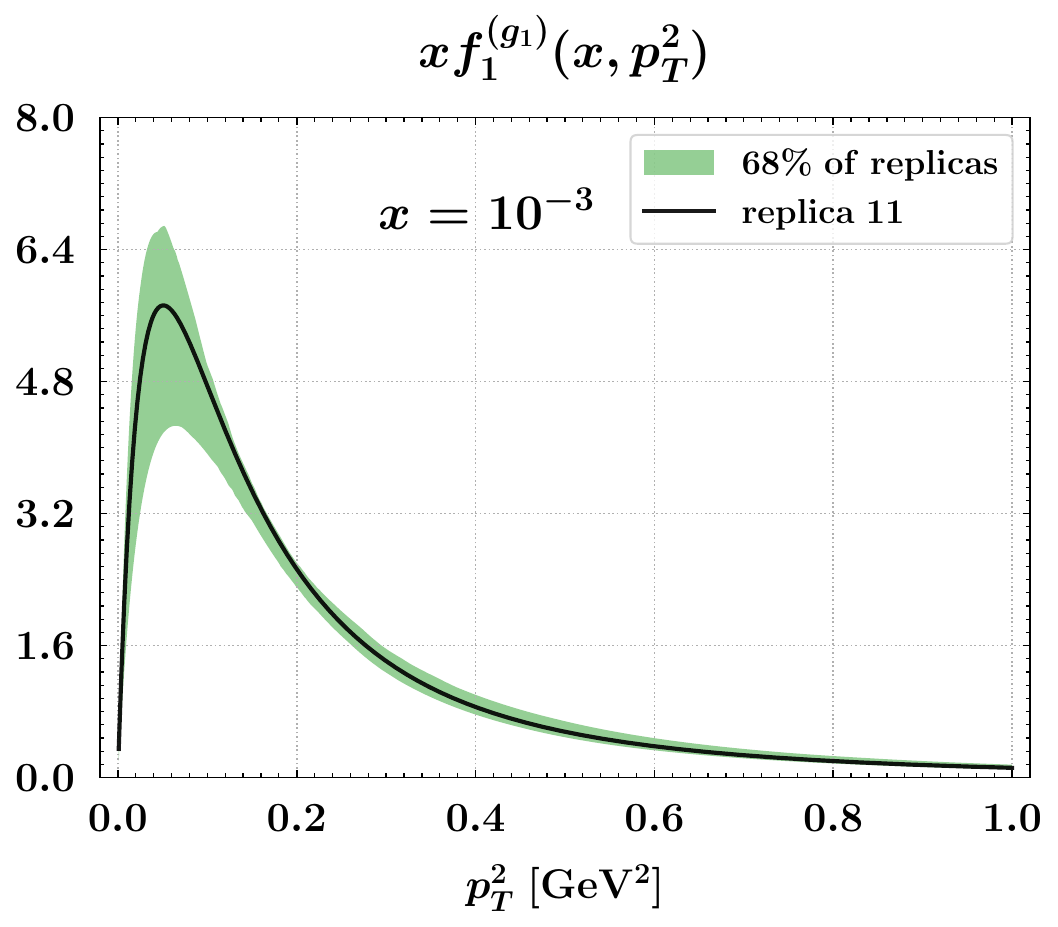}
 \hspace{0.695cm}
 \includegraphics[width=0.47\textwidth]{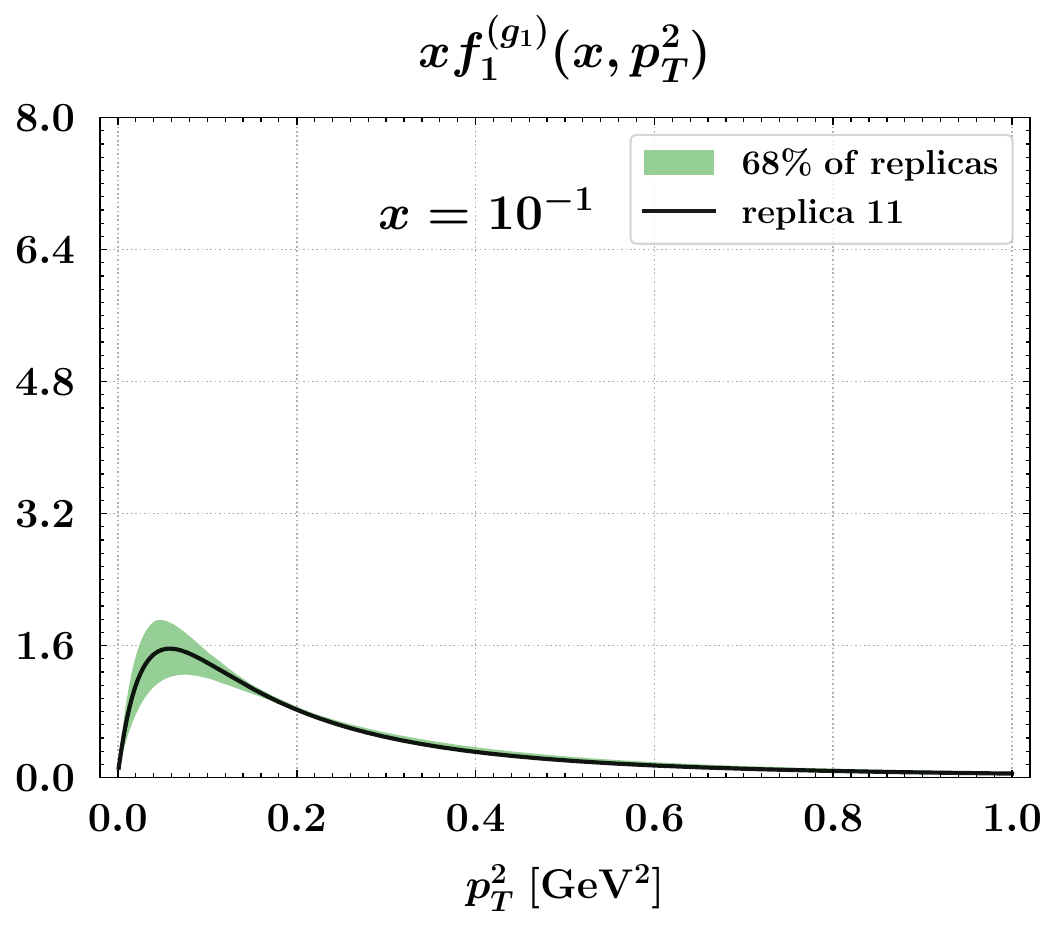}

 \vspace{0.50cm}

 \includegraphics[width=0.48\textwidth]{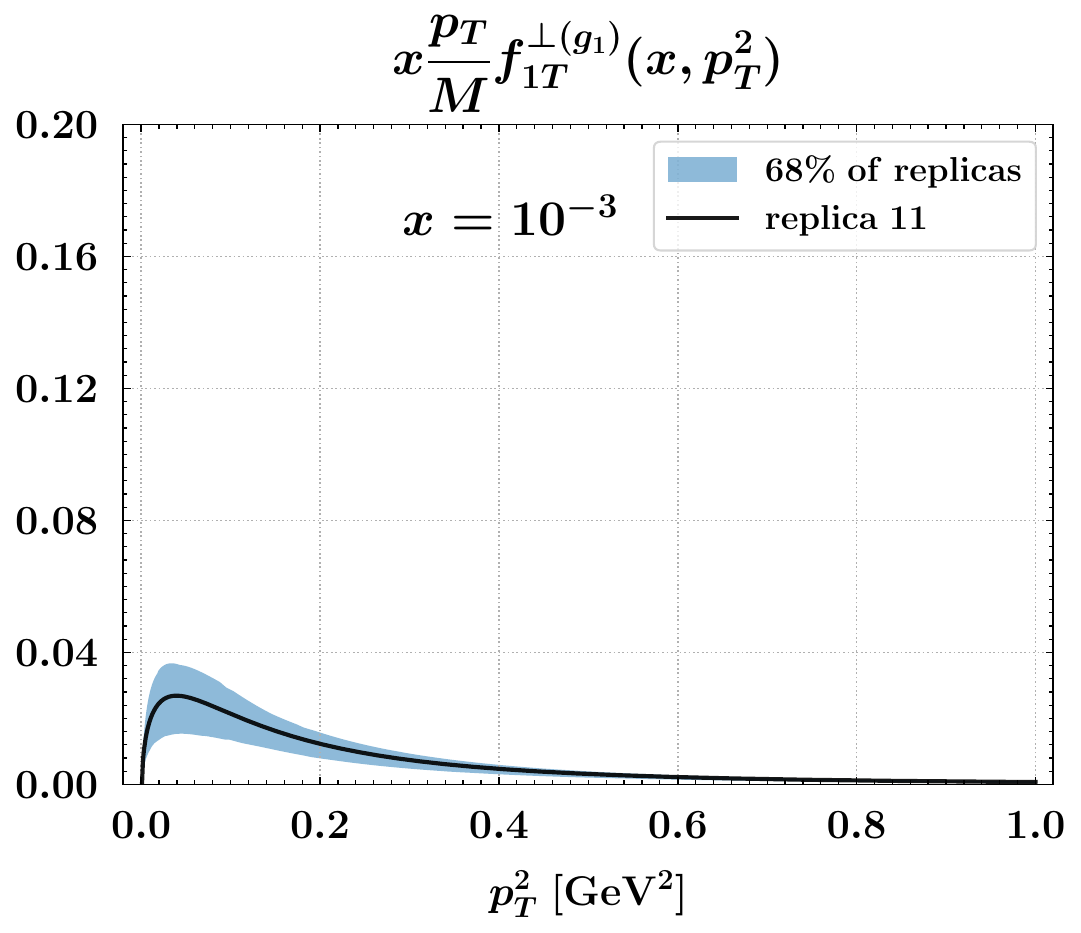}
 \hspace{0.50cm}
 \includegraphics[width=0.48\textwidth]{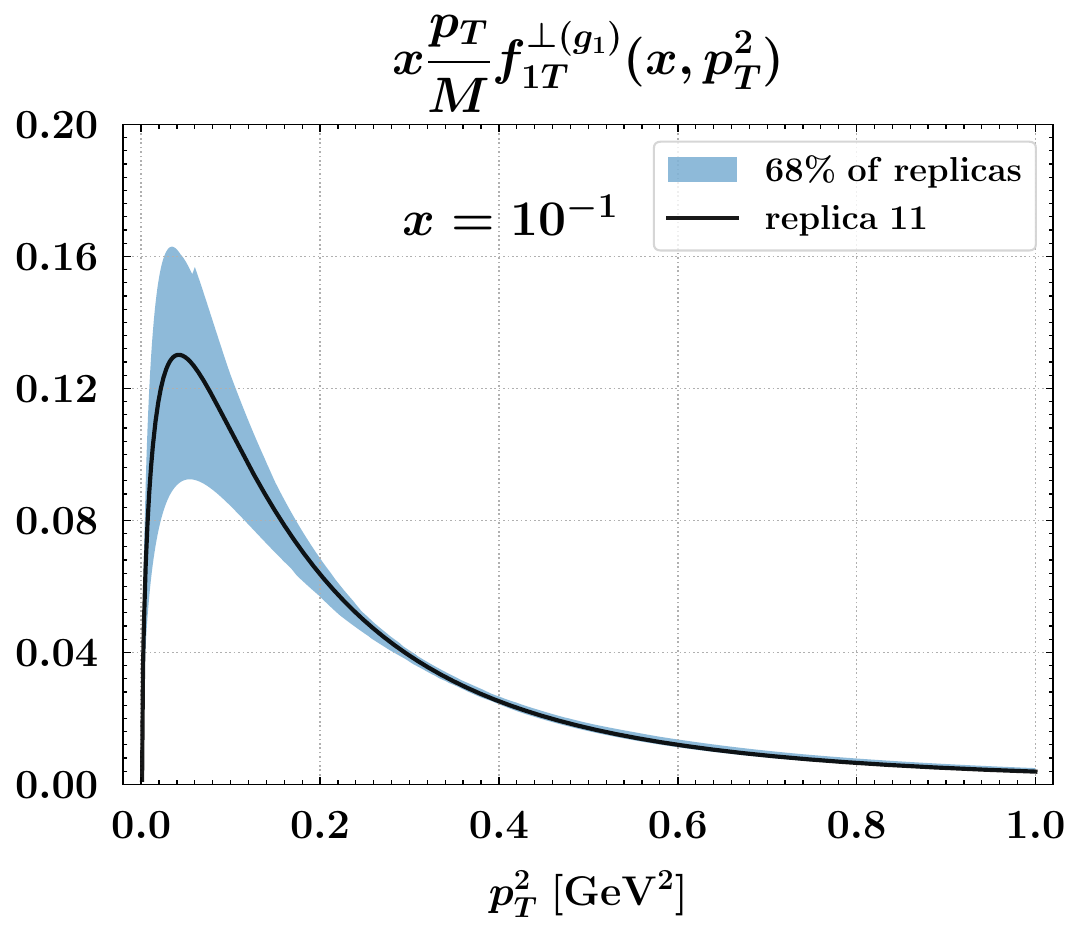}

 \caption{Transverse-momentum dependence of the $[+,+]$ unpolarized TMD (upper panel) and gluon Sivers TMD multiplied by $x |\pT|/M$ (lower panel) calculated in the $g_1$-vertex approximation (see text), at $x=10^{-3}$ (left panels) and $x=10^{-1}$ (right panels) and at the initial scale $Q_0 = 1.64$~GeV. Uncertainty band by including the 68\% of all computed replicas. Black curves refer to the most representative replica 11 (see text). 
 }
\label{fig:unpol_Sivers_g1V}
\end{figure}

In Fig.~\ref{fig:unpol_Sivers_g1V}, we show the results for the $f$-type unpolarized gluon TMD (upper panels) and the gluon Sivers function multiplied by $x |\pT| / M$ (lower panels), as functions of $\pT^2$, in the $g_1$-vertex approximation.~\footnote{Preliminary results for the $f$-type Sivers function in the $g_1$-vertex approximation were previously presented in Refs.~\cite{Bacchetta:2021lvw,Bacchetta:2022esb}.}
Left (right) plots are for TMDs calculated at $x=10^{-3}$ ($x=10^{-1}$) and at $Q_0 = 1.64$ GeV. As for the parameter values, the 68\% uncertainty bands are formed by excluding the largest and smallest 16\% of 100 computed replicas. The black solid line is the result of the most representative replica 11. Here, and in the following, the strong coupling constant is fixed to $g_s = \sqrt{\alpha_s (Q_0)} = 0.57583$. 
The qualitative behavior of the TMD $f_1^g$ 
stays practically the same with respect to the original fit (see upper panels of Fig.~4 in Ref.~\cite{Bacchetta:2020vty}). 
The resulting gluon Sivers function 
decreases at low $x$. 
However, 
this trend can radically change when including also the $g_2$ vertex,  
as shown in Sec.~\ref{s:TMDs_full}.

 \begin{table}
 \begin{tabular}[c]{|c||r|r||r|r|}
 \hline \hline
 \multicolumn{1}{l}{} & \multicolumn{2}{c}{{\tt PVGlue20}} & \multicolumn{2}{c}{{\tt PVGlue20g1V}} \\
 \hline
 parameter            &    mean           & replica 11 &    mean           & replica 11 \\
 \hline			
 $A$ [GeV$^{-1}$]     & 6.1 $\pm$ 2.3     & 6.0        & 4.3 $\pm$ 1.5     & 4.29       \\
 $a$                  & 0.82 $\pm$ 0.21   & 0.78       & 0.73 $\pm$ 0.14   & 0.73       \\
 $b$                  & 1.43 $\pm$ 0.23   & 1.38       & 1.34 $\pm$ 0.13   & 1.33       \\
 $C$ [GeV$^{-1}$]     & 371 $\pm$ 58      & 346        & 349 $\pm$ 24      & 350        \\
 $D$ [GeV]            & 0.548 $\pm$ 0.081 & 0.548      & 0.595 $\pm$ 0.049 & 0.586      \\
 $\sigma$ [GeV]       & 0.52 $\pm$ 0.14   & 0.50       & 0.42 $\pm$ 0.08   & 0.41       \\
 $\Lambda_X$ [GeV]    & 0.472 $\pm$ 0.058 & 0.448      & 0.398 $\pm$ 0.035 & 0.384      \\
 $\kappa_1$ [GeV$^2$] & 1.51 $\pm$ 0.16   & 1.46       & 1.33 $\pm$ 0.08   & 1.28       \\
 $\kappa_2$ [GeV$^2$] & 0.414 $\pm$ 0.036 & 0.414      & 0.0               & 0.0        \\
 \hline \hline
  \end{tabular}
 \caption{Mean values of fitted parameters with their 68\% uncertainties, and corresponding values for the most representative replica 11 (see text). The original fit of Ref.~\cite{Bacchetta:2020vty} and the ``$g_1$-vertex approximation" are labeled as {\tt PVGlue20} and {\tt PVGlue20g1V}, respectively (see text).}
 \label{tab:par-g}
 \end{table}

\subsection{Sivers function: Quark-target model}
\label{ss:Sivers_f_qtm}

In a similar way, we can also derive the results for 
the $f$-type gluon Sivers function in the quark-target model. We indicate it with the superscript $(q)$. 
In this approximation, the incoming proton in Fig.~\ref{fig:1-eik_Higgs_WW} is replaced by a quark, and similarly for the spectator remnant. Therefore, both the proton and the spectator mass, $M$ and $M_X$, are set equal to the target-quark mass, $m_q$, and the effective nucleon-gluon-spectator vertex is replaced by a QCD quark-gluon-quark one. Starting from the expression for the $[+,+]$ gluon correlator in the quark-target model, we obtain 
\begin{equation}
 f_{1T}^{\perp \, (q)}(x,\pT) = - \frac{4\,g_s^4}{(2\pi)^3} \frac{m_q^2\,x\,(1-x)\,P^+}{(m_q^2 x^2 + \pT^2)} \, {\cal D}_q (p) \;,
\label{f1Tp_pp_qtm_4}
\end{equation}
where
\begin{equation}
 {\cal D}_q (p) = \frac{1}{2(1-x)P^+} \int \frac{d^2 \lT}{(2\pi)^2} \frac{\lT \cdot \pT}{\pT^2}\frac{1}{\lT^2} \frac{(1-x)}{(\lT + \pT)^2 + x^2m_q^2} \;.
\label{Dq_2}
\end{equation}
Following similar steps as in the previous case, we obtain
\begin{equation}
{\cal D}_q (p) = - \frac{1}{8 \pi P^+} \frac{1}{\pT^2} \ln \frac{\pT^2 + x^2m_q^2}{x^2m_q^2} \;,
\label{Dq_4}
\end{equation}
and we get the final expression for our $f$-type gluon Sivers function in the quark-target model
\begin{equation}
 f_{1T}^{\perp \, (q)}(x,\pT) = \frac{g_s^4}{(2\pi)^4} \frac{m_q^2\,x\,(1-x)}{\pT^2 (m_q^2 x^2 + \pT^2)} \ln \frac{\pT^2 + x^2m_q^2}{x^2m_q^2} \;,
\label{f1Tp_pp_qtm}
\end{equation}
which corresponds to Eq.~(B12) of Ref.~\cite{Meissner:2007rx}.

\subsection{Linearity function: $g_1$-vertex approximation}
\label{ss:linearity_f_g1}


Let us consider now the distribution of linearly polarized gluons in a transversely polarized target, denoted as $h_1$ in Tab.~\ref{tab:gluon_TMDs}. For simplicity, we will call it ``linearity function'' even if this terminology could be used for any $h$ functions in the rightmost column of Tab.~\ref{tab:gluon_TMDs}. In spite of the similarity in notation, this function should not be confused with the analogue of the quark transversity distribution. In fact, it does not survive transverse-momentum integration and is T-odd. 

In the following, we derive the $f$-type gluon linearity function in the $g_1$-vertex approximation of our spectator model. Using the corresponding projector from Eq.~\eqref{eq:projectors_LIN}, the dipolar form for $g_1(p^2)$ as in Eq.~\eqref{eq:dipolarff}, and the $[+,+]$ gluon correlator of Eq.~\eqref{e:Phi++}, we have
\label{h1_pp_qtm_1}
\begin{equation}
\begin{split}
 h_{1}^{(g_1)}(x,\pT) &= \mathbb{P}_{[h_1^{g}]}^{\mu \nu} \, \left[ \Phi_{\mu\nu}^{[+,+]\, (g_1)} (x, \pT, \ST) - \Phi_{\mu\nu}^{[+,+] \, (g_1)} (x, \pT, -\ST) \right] 
 \\ 
 &= \frac{96\,g_s \kappa_1^3}{(2\pi)^3} \frac{M \left[ M - M_X (1-x) \right]\,P^+}{(1-x)\,\pT^2\,(p^2 - \Lambda_X^2)^2} 
 \\ 
 &\qquad \times  2 \, {\rm Re} \int \frac{d^4 l}{(2\pi)^4} \frac{l_T \cdot p_T}{(l^2 - \Lambda_X^2)^2} \frac{1}{l^+ + i\epsilon} \frac{1}{(l + p - P)^2 - M_X^2 + i\epsilon} \frac{1}{[(p + l)^2 - \Lambda_X^2 + i\epsilon]^2} \;.
\label{h1_pp_g1_1}
\end{split}
\end{equation}
Following the same steps described in  Section~\ref{ss:Sivers_f_g1} we obtain
\begin{equation}
 h_{1}^{(g_1)}(x,\pT) = - \frac{96\,g_s \kappa_1^3}{(2\pi)^3} \frac{M \left[ M - M_X (1-x) \right]\,(1-x)^4\,P^+}{[\pT^2 + L_X^2(\Lambda_X)]^2} \, {\cal D}_2 (p) \;,
\label{h1_pp_g1_2}
\end{equation}
where ${\cal D}_2 (p)$ is defined and computed in Eqs.~\eqref{D2_1}-\eqref{D2_3}.
The final expression for our $f$-type gluon linearity function in the $g_1$-vertex approximation is
\begin{equation}
\begin{split}
 h_{1}^{(g1)}&(x,\pT) = \frac{24 g_s \kappa_1^3}{(2\pi)^4} \frac{M \left[ M - M_X (1-x) \right]\,(1-x)^4}{[\pT^2 + L_X^2(\Lambda_X)]^2} \\
 &\quad \times \, \left[ \frac{1 - 2 \, \LXtwoL/\pT^2}{\LXtwoL \, [\pT^2 + 4 \LXtwoL]^2} + 8 \frac{\pT^2 + \LXtwoL}{|\pT|^3 \, [\pT^2 + 4 \LXtwoL]^{\nicefrac{5}{2}}} \, \tanh^{-1} \left( \frac{|\pT|}{\sqrt{\pT^2 + 4 \LXtwoL}} \right) \right] \;.
\label{h1_pp_g1}
\end{split}
\end{equation}

Preliminary results on the $f$-type gluon linearity function in the $g_1$-vertex approximation were presented in Refs.~\cite{Bacchetta:2021twk,Bacchetta:2022esb}.

\subsection{Linearity function: Quark-target model}
\label{ss:linearity_f_qtm}

In the quark-target model, following an analogous procedure to the one in Section~\ref{ss:Sivers_f_qtm}, we get
\begin{equation}
 h_{1}^{(q)}(x,\pT) = - \frac{8\,g_s^4}{(2\pi)^3} \frac{m_q^2\,x\,P^+}{(m_q^2 x^2 + \pT^2)} \, {\cal D}_q (p) \;.
\label{h1_pp_qtm_2}
\end{equation}
Combining Eqs.~\eqref{h1_pp_qtm_2} and~\eqref{Dq_4}, we get the final expression for our $f$-type gluon linearity function in the quark-target model
\begin{equation}
 h_{1}^{(q)}(x,\pT) = \frac{g_s^4}{(2\pi)^4} \frac{2x\,m_q^2}{\pT^2 (m_q^2 x^2 + \pT^2)} \ln \frac{\pT^2 + x^2m_q^2}{x^2m_q^2} \;,
\label{h1_pp_qtm}
\end{equation}
which corresponds to Eq.~(B17) of Ref.~\cite{Meissner:2007rx}.

\section{T-odd gluon TMDs: Results of full calculation}
\label{s:TMDs_full}

If we include the full structure of the nucleon-gluon-spectator vertex $\mathcal{Y}^{ba}_\rho$ in Eq.~\eqref{eq:vertex_ngs}, 
a given T-odd gluon TMD, generically indicated by $F(x, \pT^2)$, can be organized as 
\begin{equation}
\label{eq:F_lin_comb}
 F(x,\pT^2) = \sum_{i,j,k}^{1,2} C_{ijk}^{[F]}(x,\pT^2) \, g_s \, \kappa_i \, \kappa_j \, \kappa_k \,,
\end{equation}
where $\kappa_{i,j,k}$ are the coupling constants encoded in the dipolar form factors of  Eq.~\eqref{eq:dipolarff} with the assumption made in Eq.~\eqref{eq:g}, and $C_{ijk}^{[F]}$ are related coefficients.
For each T-odd gluon TMD $F(x, \pT^2)$, the $C_{ijk}^{[F]}$ can be split in eight different contributions $C_{ijk}^{[F], l}, \, l=1,..,8$, and organized as linear combinations according to
\begin{equation}
\label{eq:C_lin_comb}
 C_{ijk}^{[F]}(x,\pT^2) = \frac{(1-x)^4 P^+}{{(2\pi)^3} \, [\pT^2 + L_X^2(\Lambda_X)]^2} \sum_{l=1}^8 {\cal C}_{ijk}^{[F],l}(x,\pT^2) \, {\cal D}_l(x,\pT^2) \; ,
\end{equation}
where ${\cal D}_l(x,\pT^2)$ are eight different master integrals that can be found in Appendix~\ref{a:integrals}. The final expressions of the $C_{ijk}^{[F], l}$ coefficients for each T-odd gluon TMD $F$ and for $l=1,..,8$ and $i,j,k=1,2$, are listed in Appendix~\ref{a:coeffs}.

We note that both the T-odd $f$-type $h_{1L}^{\perp}$ and $h_{1T}^{\perp}$ vanish in the $g_1$-vertex approximation and in the quark-target model, because the integral describing the loop in Fig.~\ref{fig:1-eik_Higgs_WW} would be proportional to $l^+$, which is set to zero by the first 
of the two Cutkosky rules 
in Eq.~\eqref{prop_delta_f1Tp_pp}. This result is in line with Eqs.~(B16) and (B18) of Ref.~\cite{Meissner:2007rx}, respectively.




In the following, we show the results of the full calculation of all the four T-odd $f$-type gluon TMDs that appear at leading twist (see Tab.~\ref{tab:gluon_TMDs}). We recall that in our model T-odd $d$-type gluon TMDs turn out to be equal to the $f$-type ones up to a color factor computed in Eq.~\eqref{e:color_factor}, because in the vertices we take the same dipole-like couplings $g_{1,2}(p^2)$ for $f$-type and $d$-type functions. Moreover, the parameters of both T-odd $f$-type and $d$-type functions are fully determined by those ones entering the T-even gluon TMDs. These parameters were fixed in Ref.~\cite{Bacchetta:2020vty} by fitting the integrated T-even gluon TMDs onto the corresponding known collinear PDFs at the low scale $Q_0 = 1.64$ GeV; their values are listed in the columns of Tab.~\ref{tab:par-g} labelled by {\tt PVGlue20}. 

It is convenient to start from the $f$-type gluon Sivers function $f_{1T}^{\perp}$ in order to compare with the results displayed in the lower panels of Fig.~\ref{fig:unpol_Sivers_g1V} using the $g_1$-vertex approximation. 



\begin{figure}[t]
 
 \centering

 \includegraphics[width=0.48\textwidth]{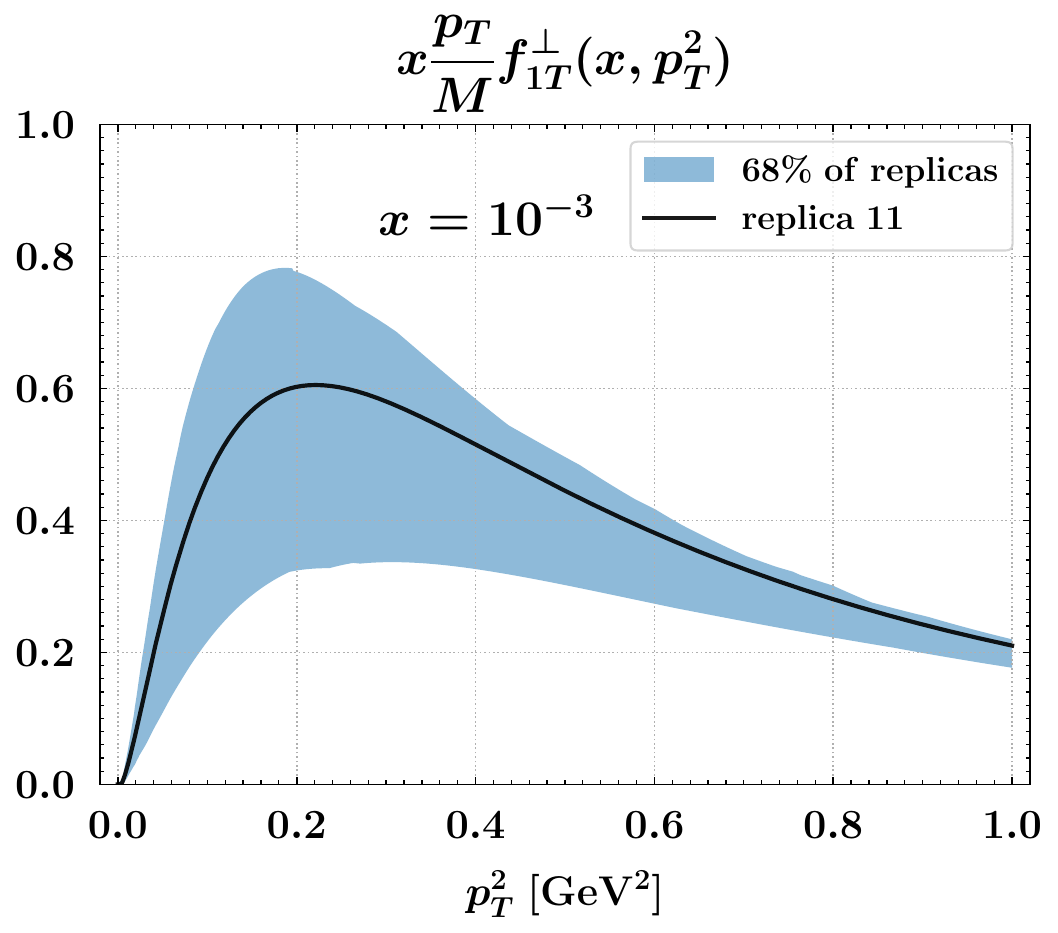}
 \hspace{0.50cm}
 \includegraphics[width=0.48\textwidth]{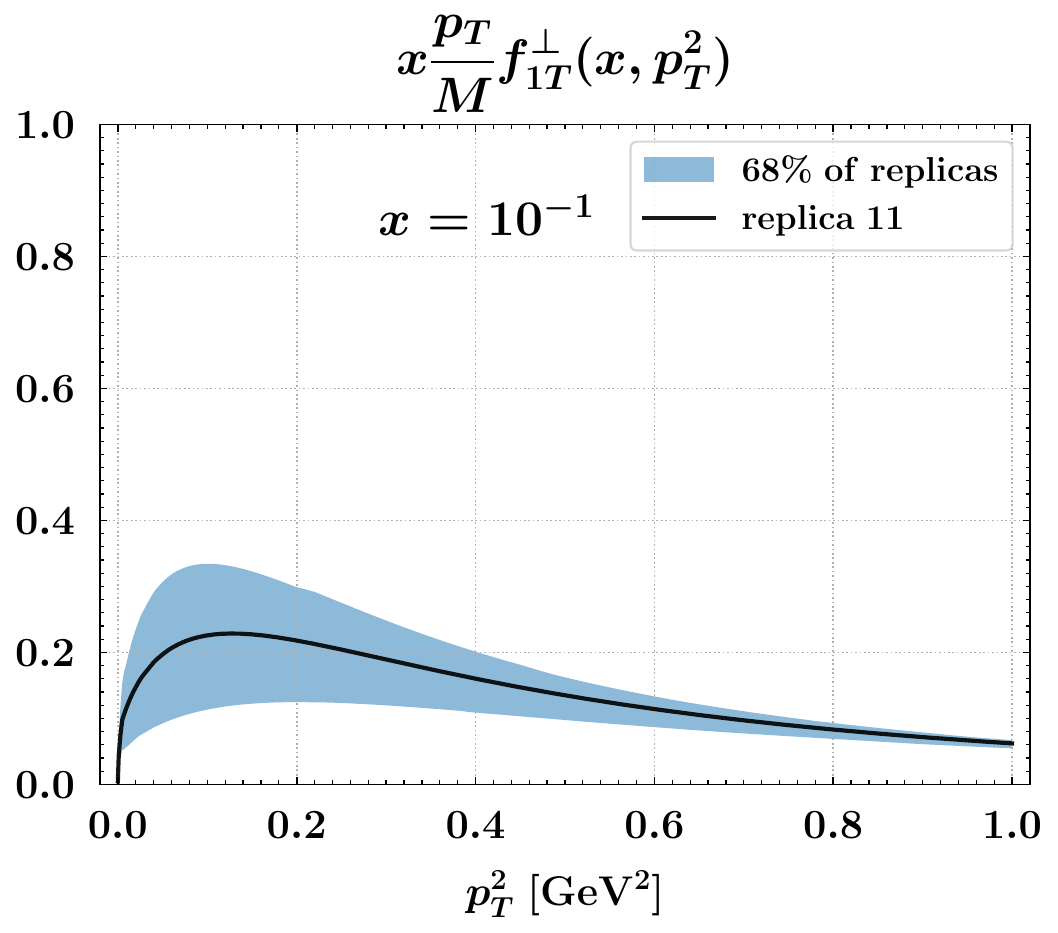}

 \vspace{0.50cm}
 \includegraphics[width=0.48\textwidth]{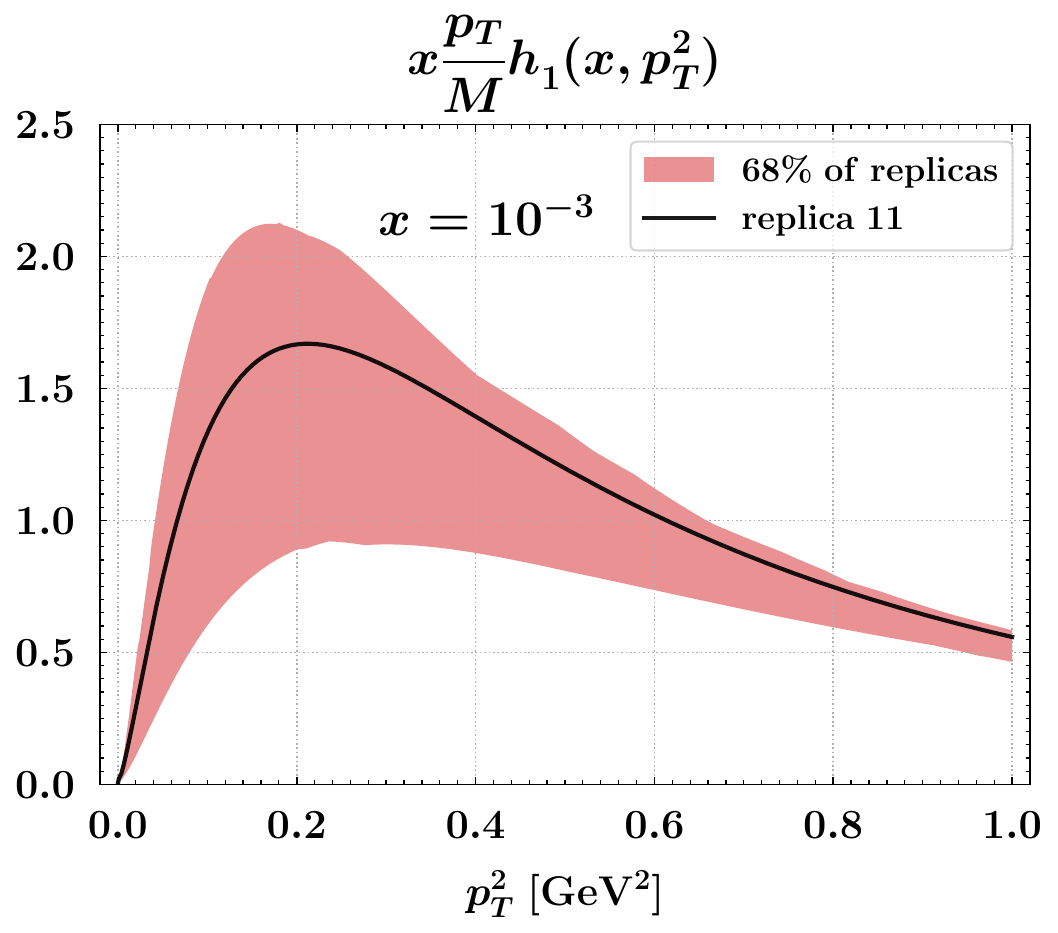}
 \hspace{0.50cm}
 \includegraphics[width=0.48\textwidth]{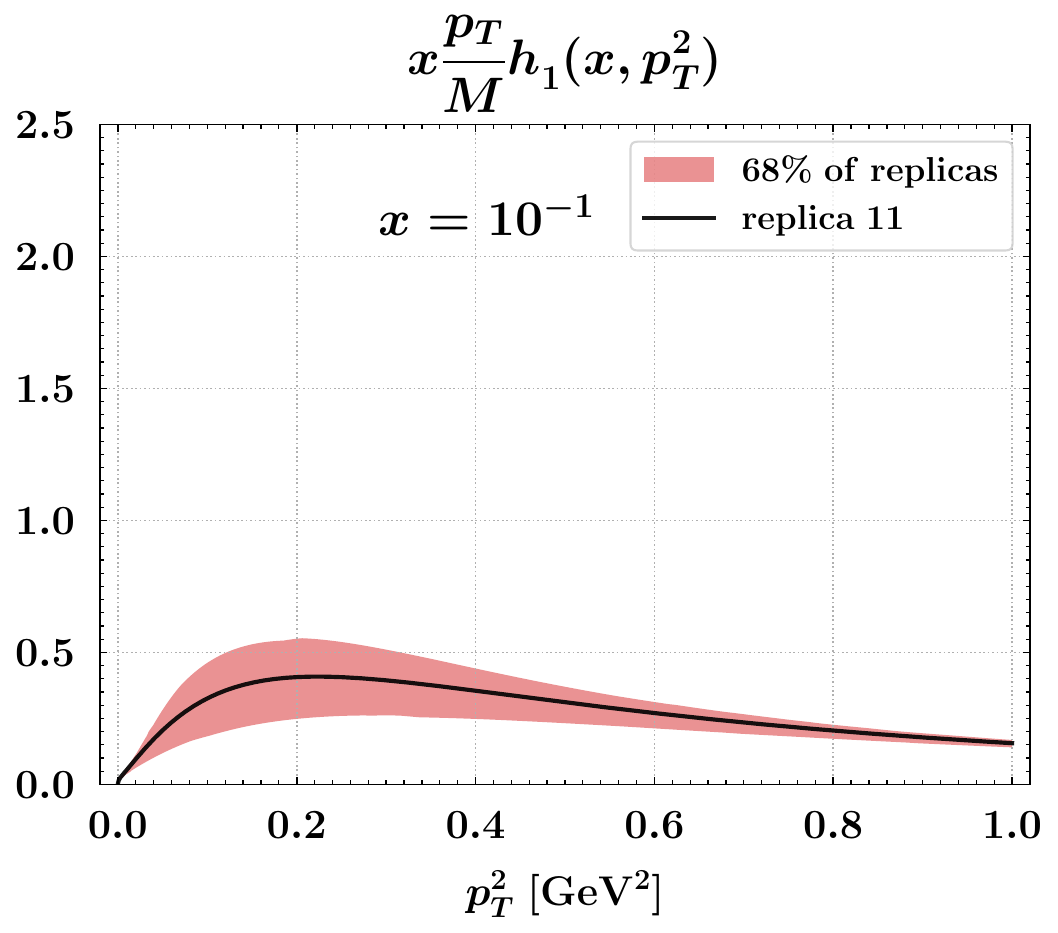}

 \caption{Transverse-momentum dependence of the $[+,+]$ gluon Sivers (upper) and linearity (lower) functions multiplied by $x |\pT|/M$, as functions of $\pT^2$ at $x=10^{-3}$ (left), and $x=10^{-1}$ (right) and at the initial scale $Q_0 = 1.64$~GeV. Uncertainty band from 68\% of all computed replicas. Black curves refer to the most representative replica 11 (see text).}
\label{fig:Sivers_linearity}
\end{figure}

In the upper panels of Fig.~\ref{fig:Sivers_linearity}, we display the T-odd $f$-type gluon Sivers function $f_{1T}^{\perp}$ multiplied by $x |\pT|/M$, as a function of $\pT^2$ at $x=10^{-3}$ (left) and $x=10^{-1}$ (right) and at the scale $Q_0 = 1.64$ GeV. As in previous figures, the uncertainty band is constructed by excluding the largest and smallest 16\% of all 100 computed replicas, roughly corresponding to $1\sigma$ standard deviation. The solid black line is the result of the most representative replica 11. 
The observed behavior in $\pT^2$ clearly does not follow a simple Gaussian pattern, rather it shows a large flattening tail for increasing $\pT^2$. The $f$-type Sivers function is regular in $\pT^2=0$, as it can be realized by inspecting the coefficients of Eqs.~\eqref{eq:F_lin_comb},\eqref{eq:C_lin_comb} listed in Tabs.~\ref{tab:SIV1}-\ref{tab:SIV4} and the master integrals in Appendix~\ref{a:integrals}. Hence, the combination $x |\pT|/M \, f_{1T}^{\perp}$ vanishes at $\pT^2=0$. 

By comparing with the lower panels of Fig.~\ref{fig:unpol_Sivers_g1V} where the $x |\pT|/M\, f_{1T}^{\perp (g_1)}$ was computed in the $g_1$-vertex approximation, we realize that the contribution of the $g_2$ coupling to the vertices $\mathcal{Y}^{ba}_\rho$ of Eq.~\eqref{eq:vertex_ngs} and $\mathcal{X}^{bde}_\alpha$ of Eq.~\eqref{eq:vertex_sgs} completely reverses the situation: the $f$-type gluon Sivers function now increases for decreasing $x$, thus supporting the statement that spin asymmetries generated by this T-odd gluon TMD could be sizable also at small-$x$.

In the lower panels of Fig.~\ref{fig:Sivers_linearity}, we show the result of the full calculation of the T-odd $f$-type gluon linearity function $h_1$ multiplied by $x |\pT|/M$, as a function of $\pT^2$ at $x=10^{-3}$ (left) and $x=10^{-1}$ (right) and at the scale $Q_0 = 1.64$ GeV. Notations are the same as in previous panels. The displayed trend is similar to the $f$-type Sivers function. Namely, the linearity increases with decreasing $x$, actually having a size larger than the Sivers function. The linearity is also regular at $\pT^2=0$, hence vanishes at this point when multiplied by $x |\pT|/M$. 

\begin{figure}[t]
 
 \centering

 \includegraphics[width=0.46\textwidth]{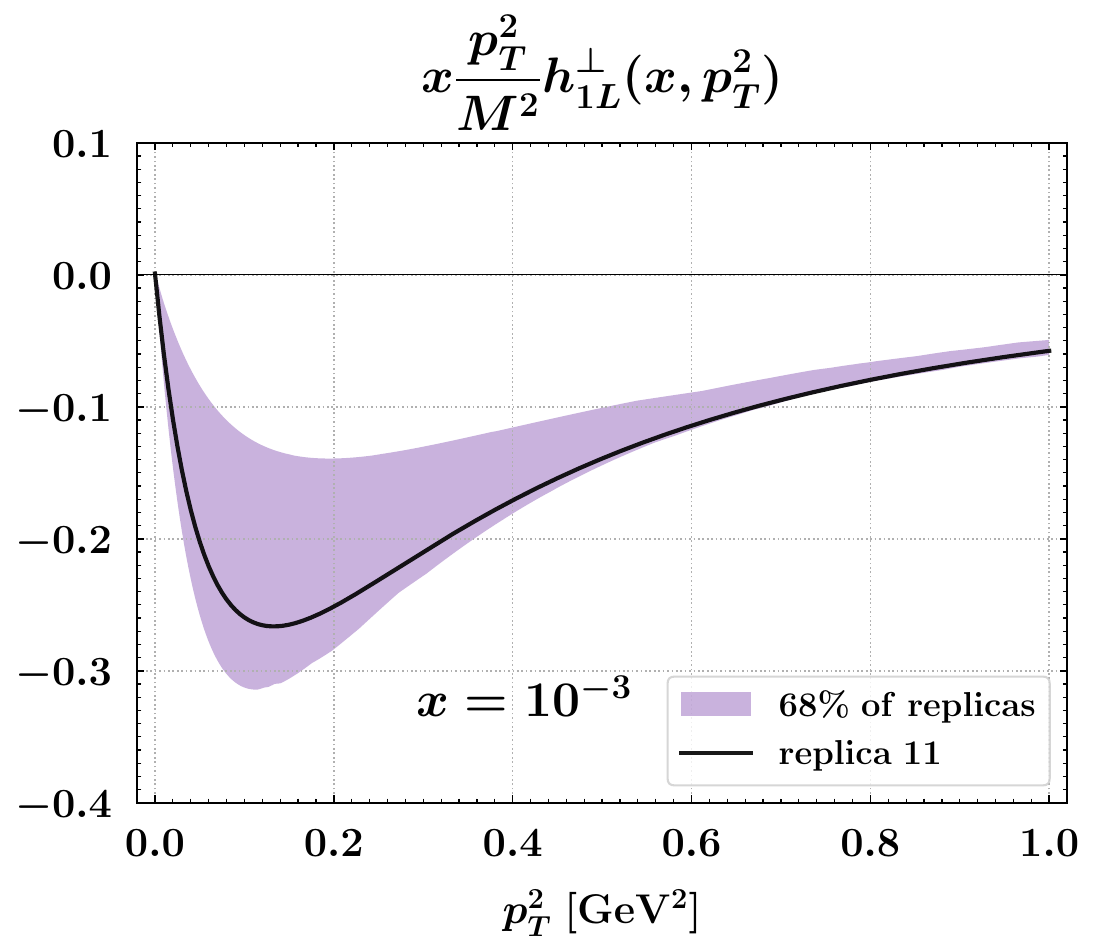}
 \hspace{0.50cm}
 \includegraphics[width=0.46\textwidth]{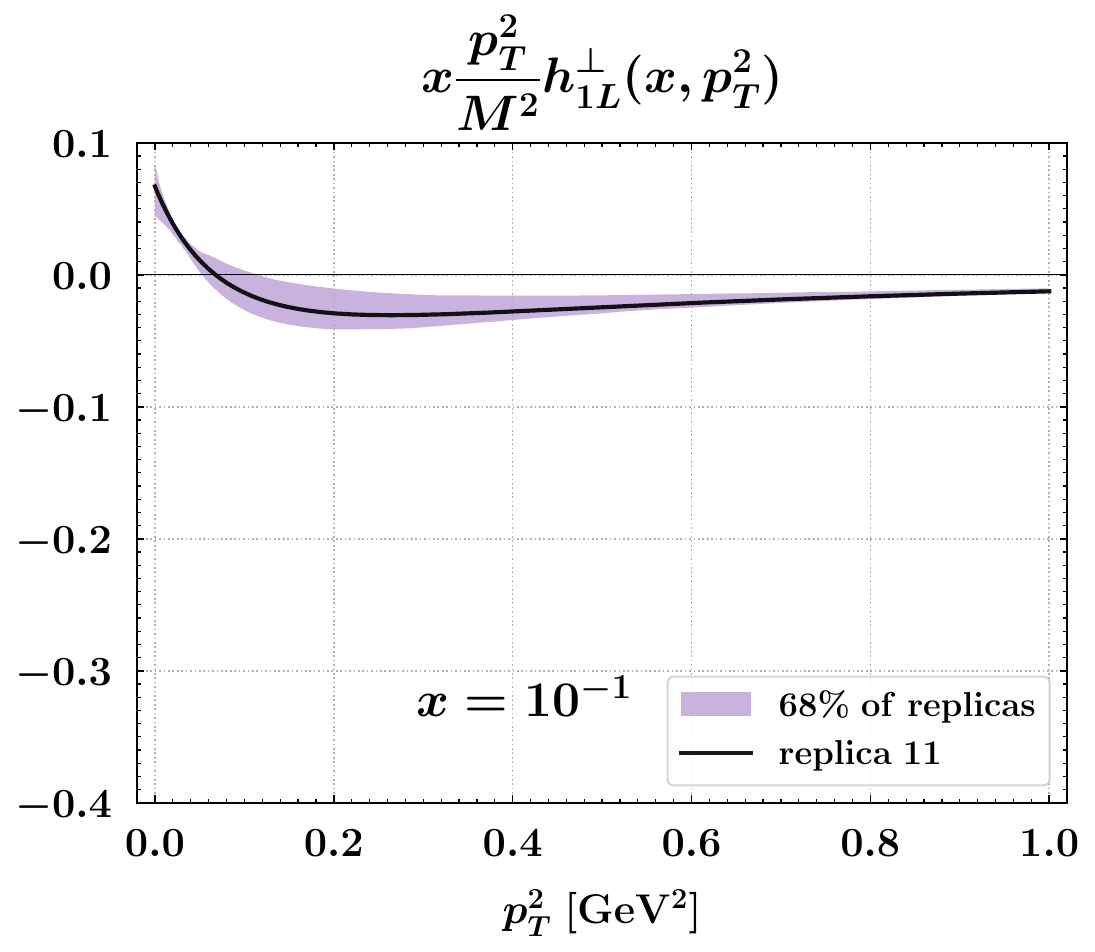}

 \vspace{0.50cm}
 \hspace{-0.50cm}
 \includegraphics[width=0.48\textwidth]{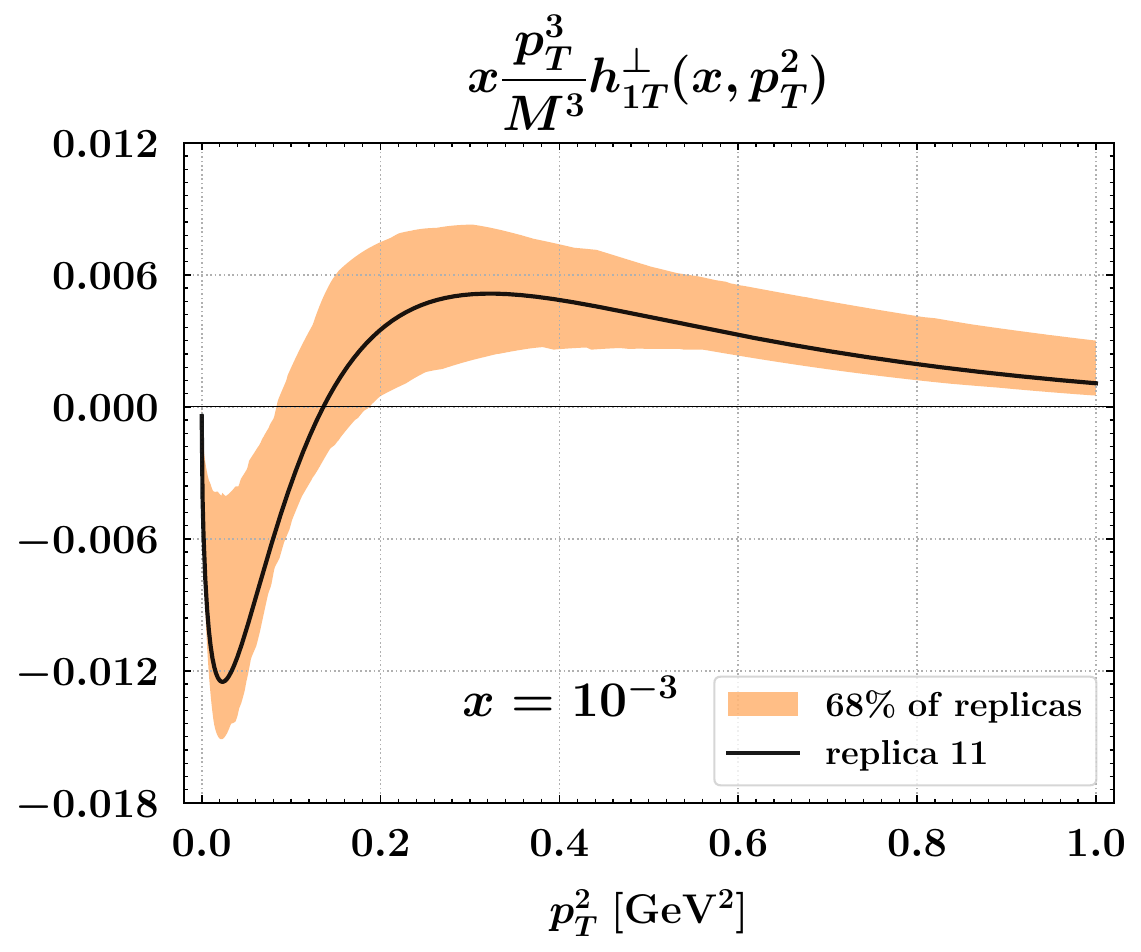}
 \hspace{0.20cm}
 \includegraphics[width=0.48\textwidth]{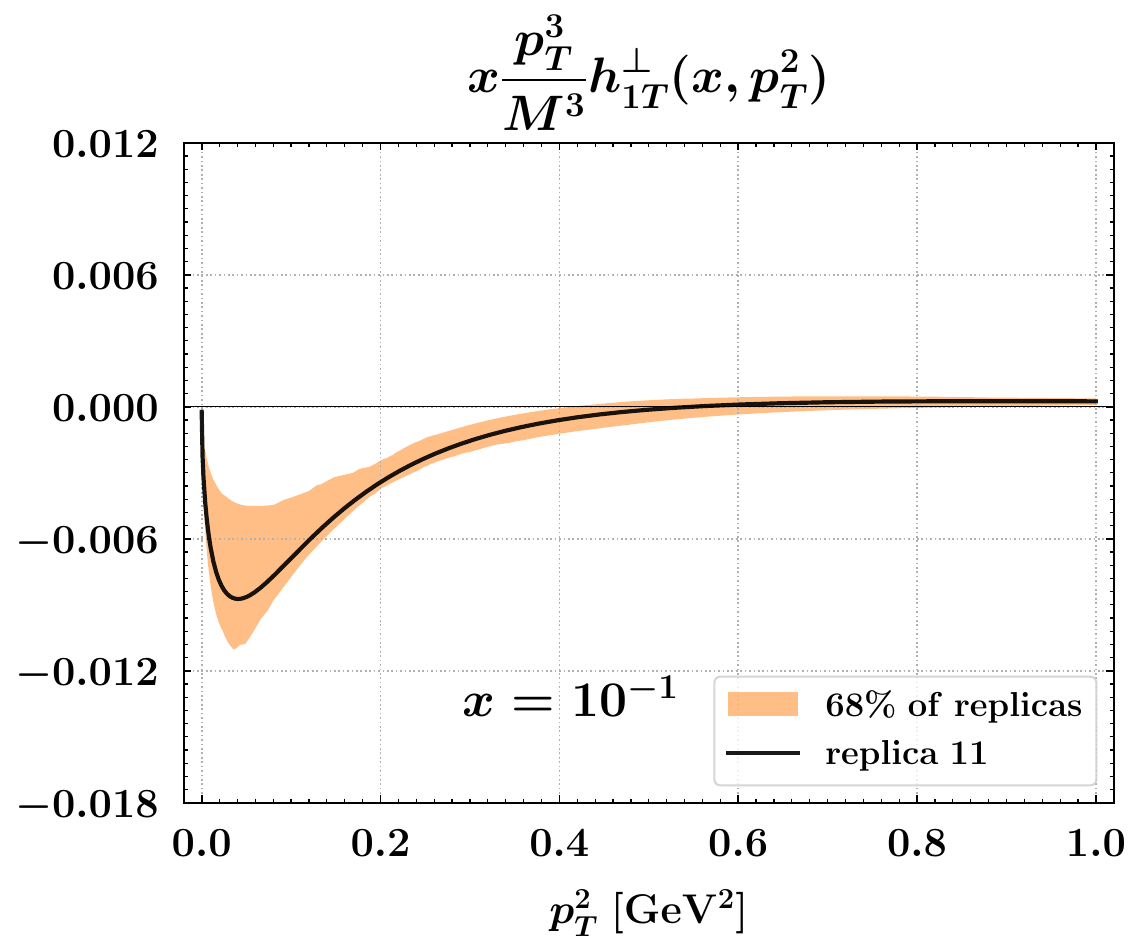}

 \caption{Transverse-momentum dependence of the $[+,+]$ gluon $h_{1L}^\perp$ (upper) and $h_{1T}^\perp$ (lower) functions multiplied by $x \pT^2/M^2$ and $x |\pT|^3/M^3$, respectively, as functions of $\pT^2$  at $x=10^{-3}$ (left) and $x=10^{-1}$ (right) and at the initial scale $Q_0 = 1.64$~GeV. Notations as in previous figures.}
\label{fig:WG_pretzelosity}
\end{figure}

In Fig.~\ref{fig:WG_pretzelosity}, we show for the first time the result of the full calculation of the T-odd $f$-type gluon $h_{1L}^\perp$ (upper panels) and $h_{1T}^\perp$ (lower panels) functions. In particular, in the upper panel we display $x \pT^2/M^2\, h_{1L}^{\perp}$ as a function of $\pT^2$ at $x=10^{-3}$ (left) and $x=10^{-1}$ (right) and at the scale $Q_0 = 1.64$~GeV. Notations are the same as in previous figures. We note that the absolute size increases with decreasing $x$, but overall it is much smaller than the Sivers and linearity functions. Interestingly, at $x=10^{-1}$ the $h_{1L}^\perp$ function shows a very long tail in $\pT^2$ but changes sign having a node at $\pT^2 \approx 0.1$ GeV$^2$. 

In the lower panels, the $x \pT^3/M^3\, h_{1T}^{\perp}$ is displayed as a function of $\pT^2$ at $x=10^{-3}$ (left) and $x=10^{-1}$ (right) and at the scale $Q_0 = 1.64$~GeV. Notations are the same as in previous figures. The absolute size is one order of magnitude smaller, raising doubts on the actual possibility of ever extracting the $h_{1T}^\perp$ from a spin asymmetry measurement. However, it shows an interesting structure with a node at small $\pT^2$ and $x$. 


\begin{figure}[t]
 
 \centering

 \hspace{0.090cm}
 \includegraphics[width=0.47\textwidth]{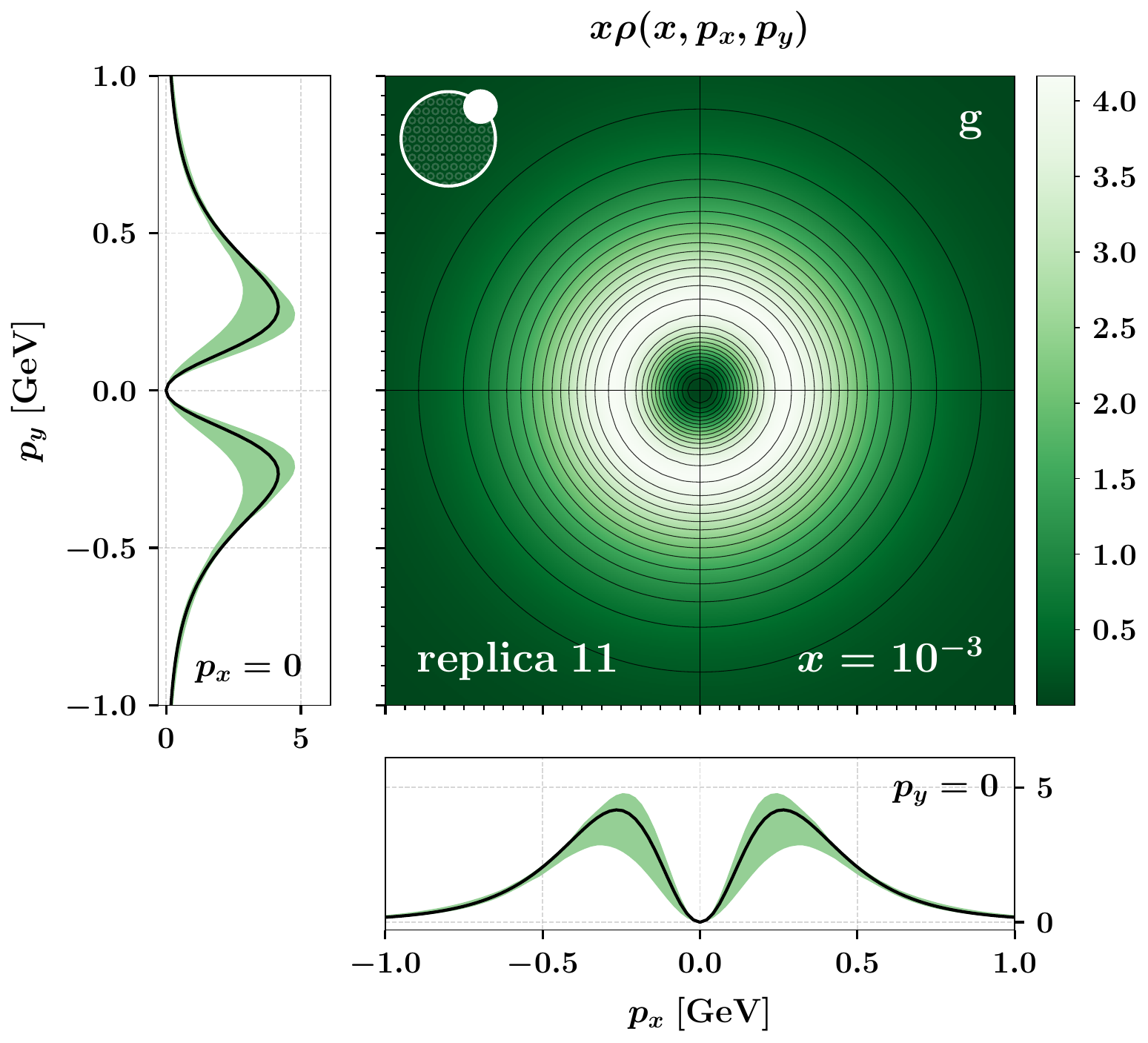}
 \hspace{0.695cm}
 \includegraphics[width=0.47\textwidth]{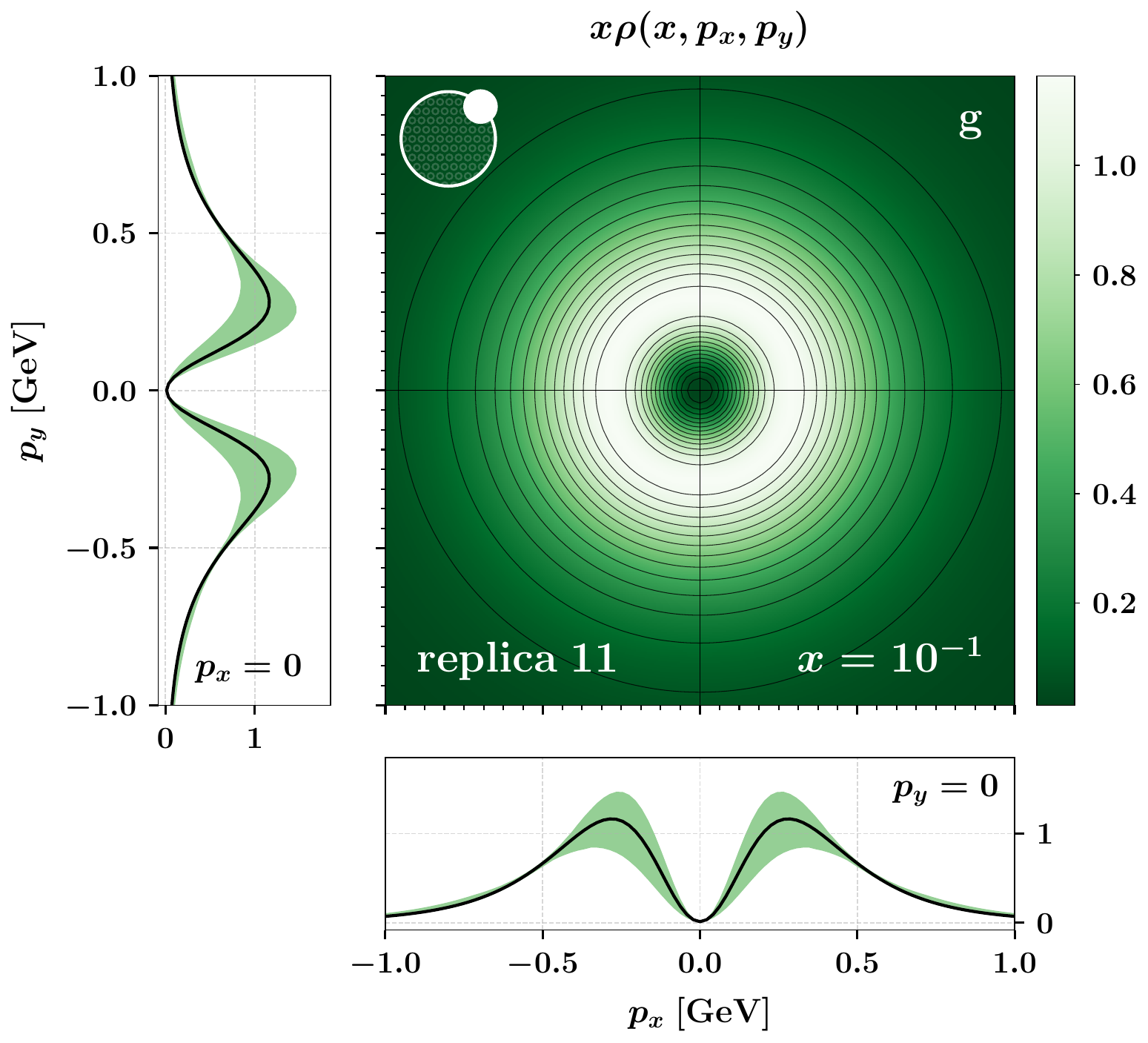}

 \vspace{0.50cm}

 \includegraphics[width=0.48\textwidth]{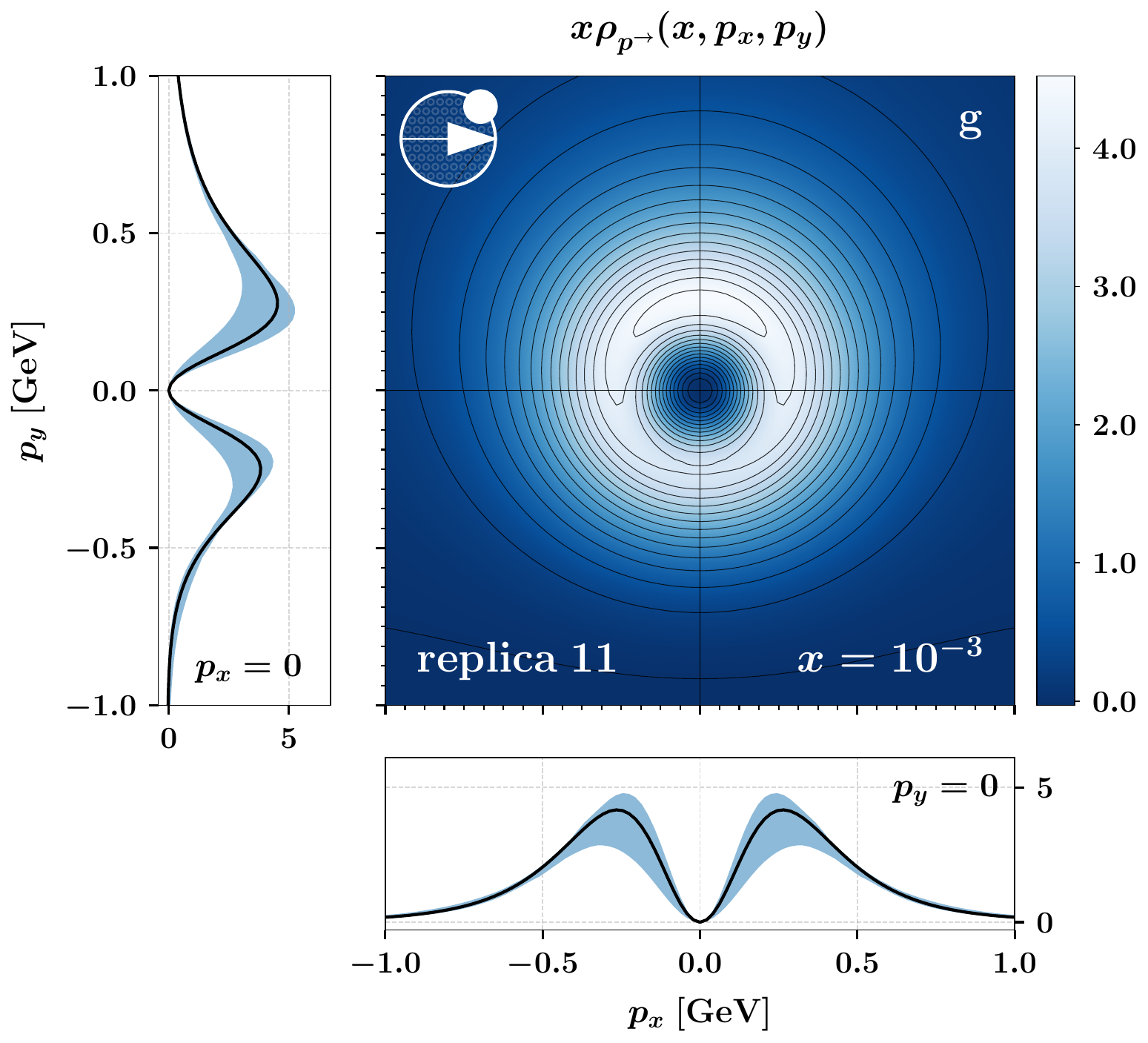}
 \hspace{0.50cm}
 \includegraphics[width=0.48\textwidth]{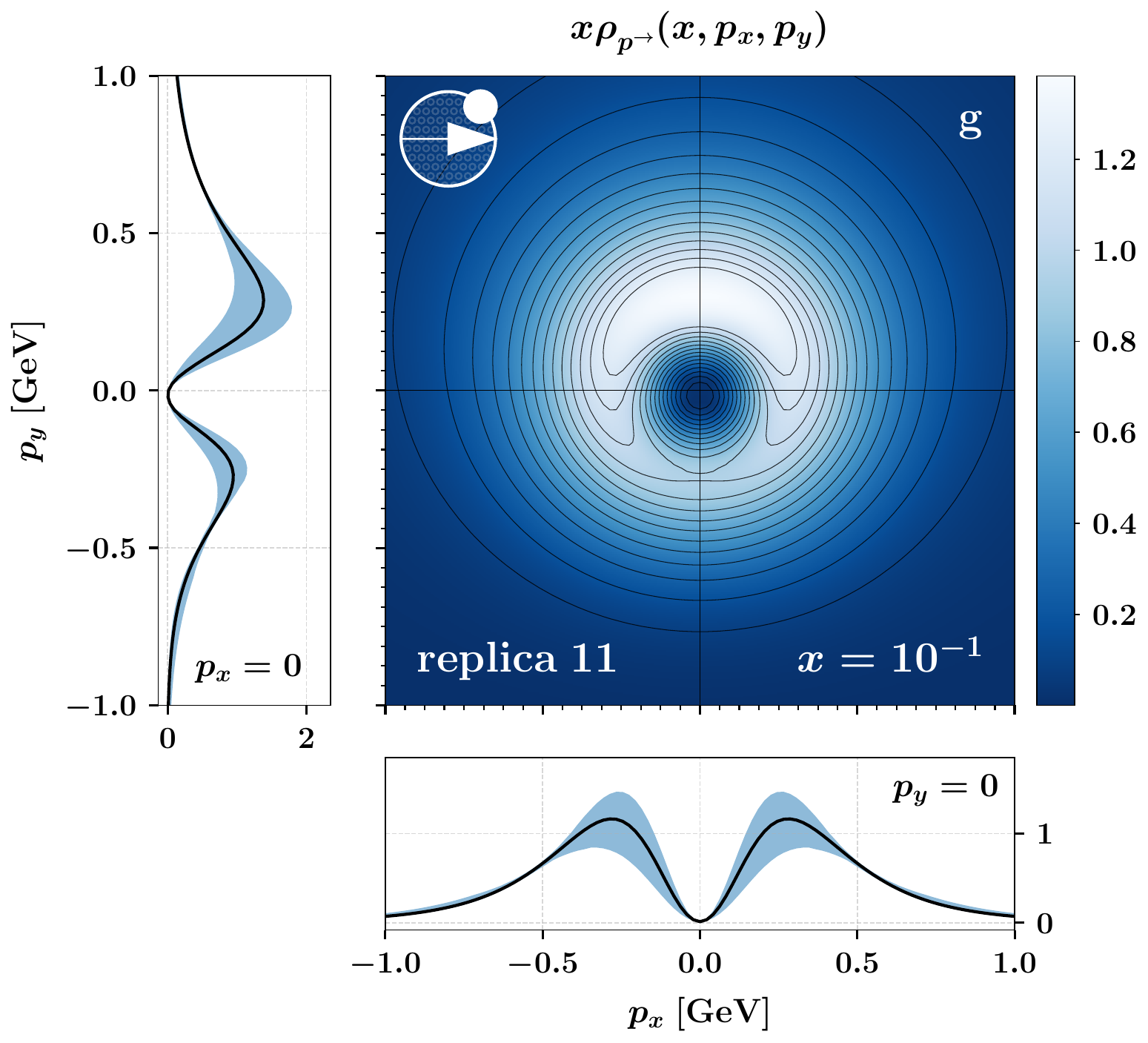}

 \caption{Unpolarized gluon density for a unpolarized (upper panels) or transvsersely polarized nucleon along $\hat{x}$ (lower panels) as a function of $\pT$ at $Q_0=1.64$ GeV and at $x=10^{-3}$ (left panels) and $x=10^{-1}$ (right panels). The nucleon is virtually moving towards the reader. Results from replica 11 (see text). Ancillary 1-dim plots for the density at $p_y=0$ and $p_x=0$ with 68\% uncertainty band. Solid black line for replica 11 (corresponding to contour plot). }
\label{fig:rho_unpol_Sivers}
\end{figure}


\begin{figure}[t]
 
 \centering

 \hspace{0.090cm}
 \includegraphics[width=0.43\textwidth]{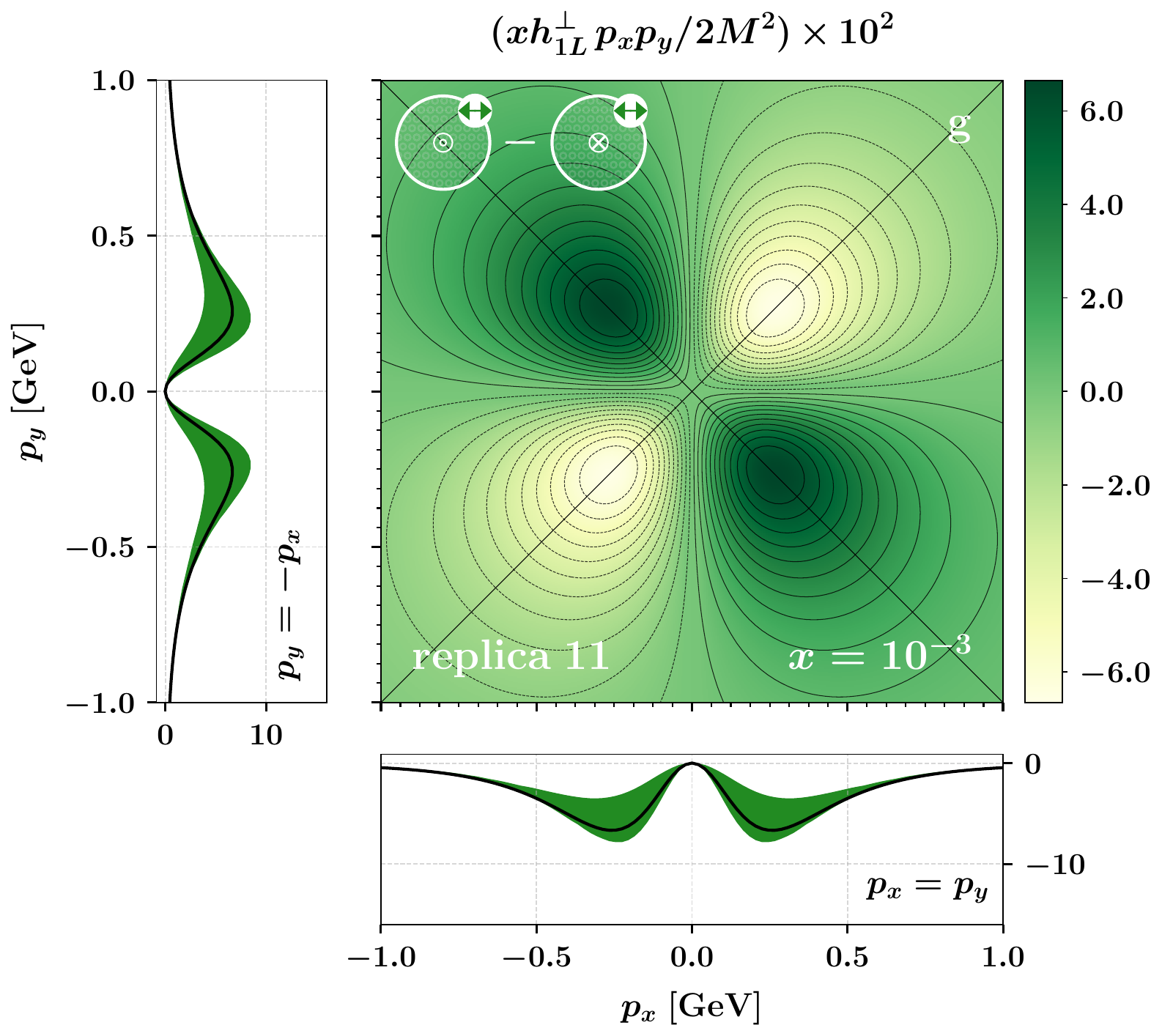}
 \hspace{0.695cm}
 \includegraphics[width=0.43\textwidth]{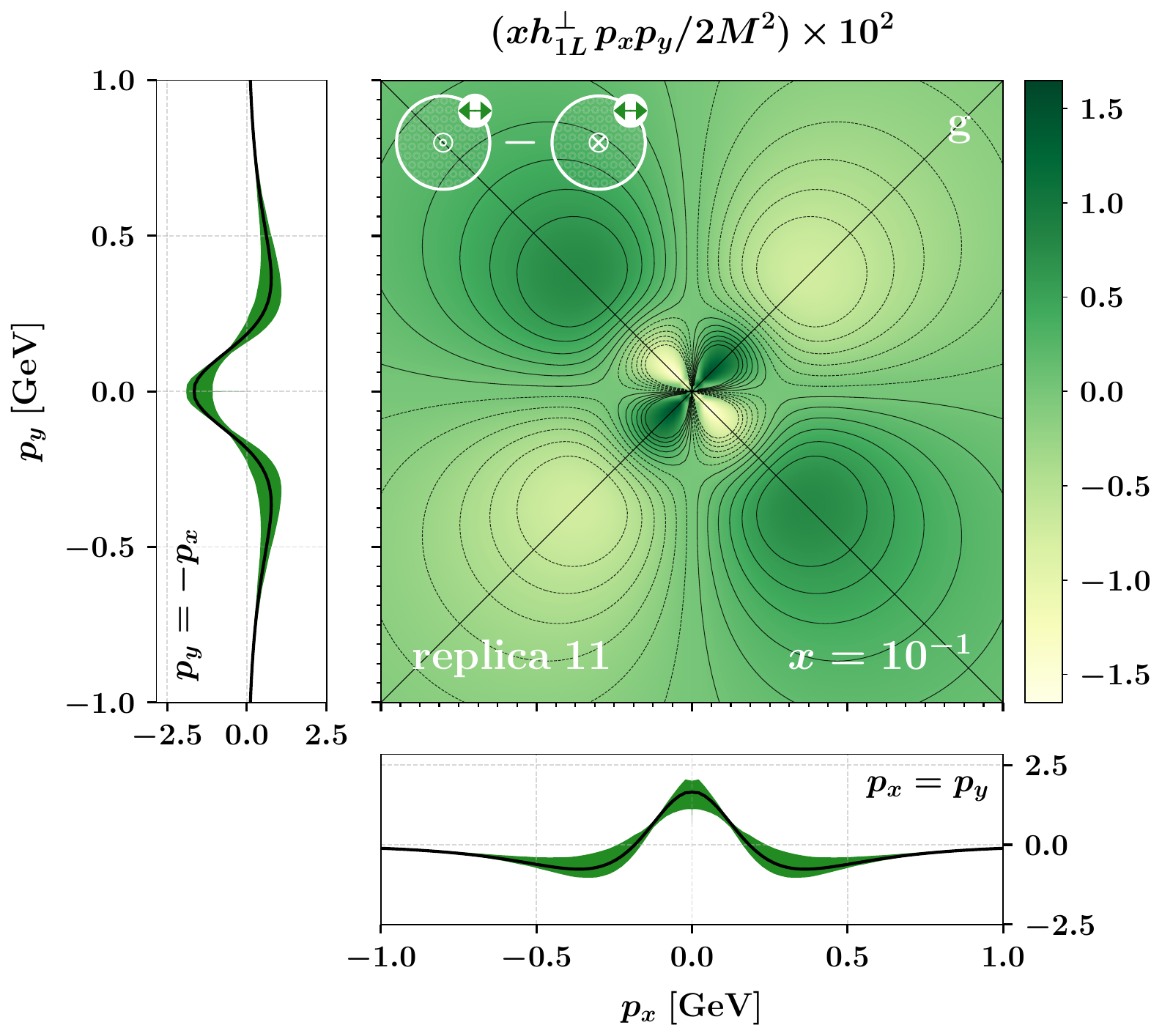}

 \vspace{0.2cm}

 \includegraphics[width=0.43\textwidth]{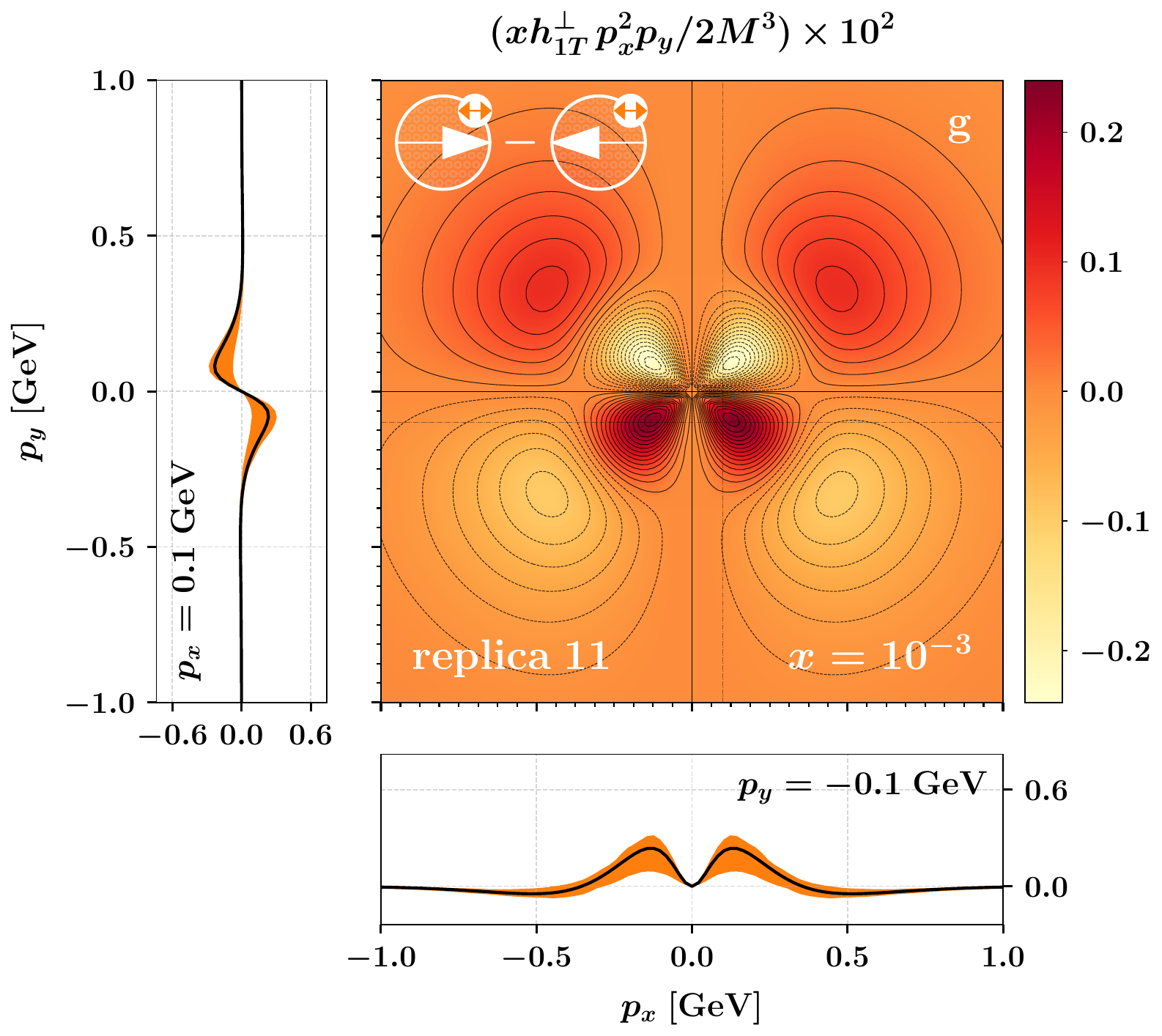}
 \hspace{0.50cm}
 \includegraphics[width=0.43\textwidth]{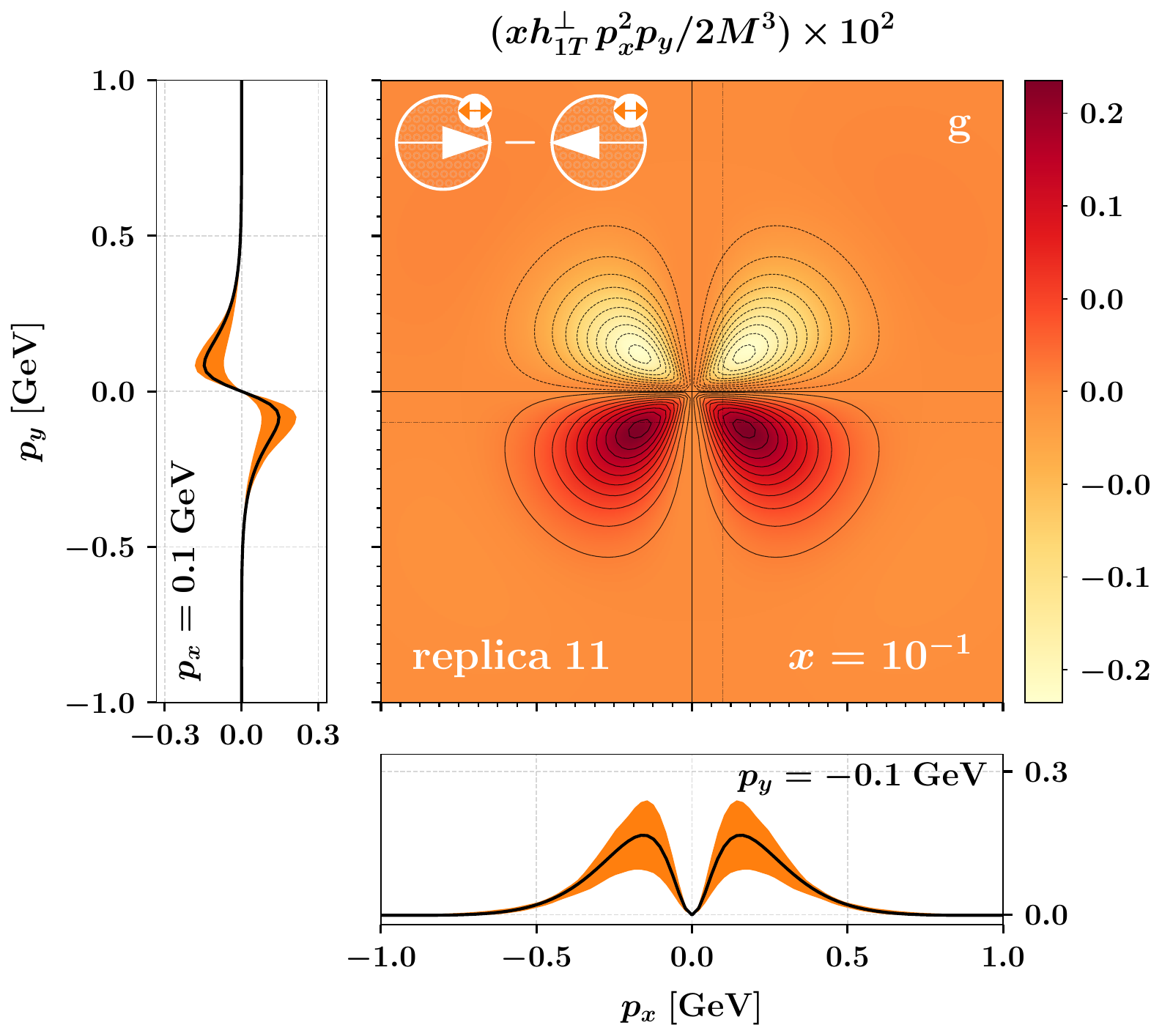}

 \vspace{0.2cm}

 \includegraphics[width=0.43\textwidth]{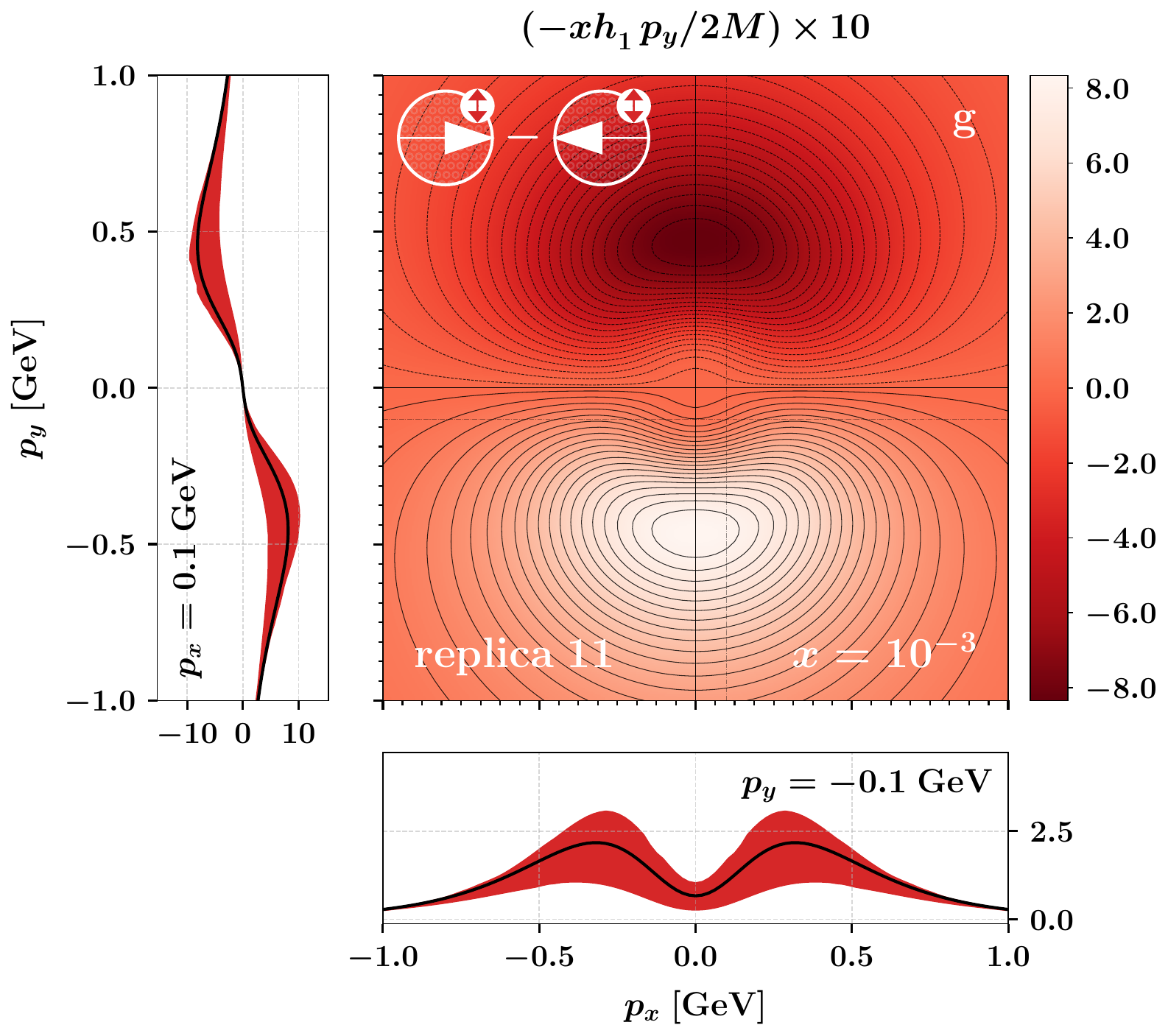}
 \hspace{0.50cm}
 \includegraphics[width=0.43\textwidth]{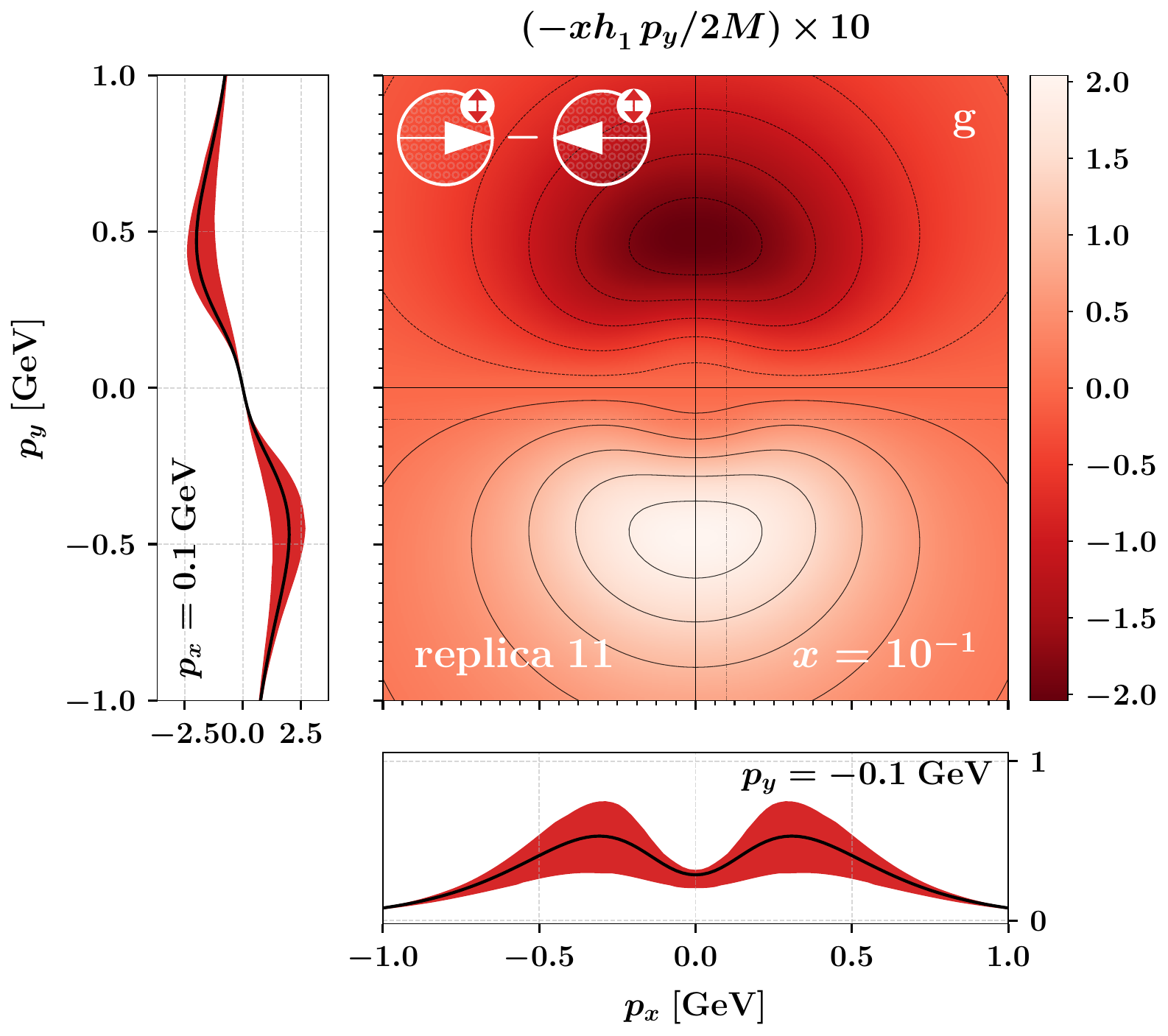}

 \caption{The 2-dim density for linearly polarized gluons in polarized nucleons as a function of $\pT$ at $Q_0=1.64$ GeV and at $x=10^{-3}$ (left panels) and $x=10^{-1}$ (right panels). The nucleon is virtually moving towards the reader. Results from replica 11 (see text). Ancillary 1-dim plots for slices of the density at specific values of $p_x$ or $p_y$, with 68\% uncertainty band and solid black line for replica 11. Upper panels: gluon linear polarization along $\hat{x}$ and nucleon longitudinal polarization, proportional to $h_{1L}^{\perp}\, p_x p_y / 2 M^2$ scaled by $10^2$. Central panels: gluon linear polarization and nucleon polarization along $\hat{x}$, proportional to $h_{1T}^{\perp}\, p_x^2 p_y / 2 M^3$ scaled by $10^2$. Lower panels: gluon linear polarization along $\hat{y}$ and nucleon polarization along $\hat{x}$, proportional to $-h_1 \, p_y/2M$ scaled by 10.}
\label{fig:rho_propeller_butterfly}
\end{figure}


Using the T-odd gluon TMDs computed in our model, we can complete the tomographic picture of the nucleon already discussed in Ref.~\cite{Bacchetta:2020vty}. To this purpose, we can construct 2-dim $\pT$-distributions of gluons at different $x$ for various combinations of their polarization and of the nucleon spin state. 

Excluding the case of a circularly polarized gluon for which no T-odd gluon TMDs occur (see Tab.~\ref{tab:gluon_TMDs}), we can have in principle six combinations: two polarization states of the gluon (unpolarized, linearly polarized) for each polarization state of the parent nucleon (unpolarized, longitudinally polarized, transversely polarized). However, the actual combinations are five, since an unpolarized gluon in a longitudinally polarized nucleon is forbidden by parity invariance (see Tab.~\ref{tab:gluon_TMDs}). 

For a unpolarized gluon in a unpolarized nucleon, we identify the 2-dim density as 
\begin{equation}
 \label{eq:rho_unpol}
 x \rho (x, p_x, p_y) = x \, f_1 (x, \pT^2)
 \; , 
\end{equation}
where $f_1$ is the leading-twist $f$-type unpolarized gluon TMD. The upper panels of Fig.~\ref{fig:rho_unpol_Sivers} show the contour plots for the $\pT$-distribution of $x \rho$ from replica 11 at $x=10^{-3}$ (left) and $x=10^{-1}$ (right) and at the scale $Q_0 = 1.64$~GeV, for a nucleon moving towards the reader. The color code identifies the size of the oscillations. For a better visualization, ancillary 1-dim plots are attached, 
representing a ``slice" of $x \rho$ at $p_x=0$ or $p_y=0$. The 68\% uncertainty band is obtained as usual by excluding the largest and smallest 16\% of 100 computed replicas; the solid black line is the result of replica 11, actually corresponding to the 2-dim contour plot.  Since both nucleon and gluon are unpolarized, the 2-dim density shows a perfect cylindrical symmetry around the direction of motion of the nucleon pointing towards the reader. 

For a unpolarized gluon in a nucleon transversely polarized along $\hat{x}$ ($|\ST|= S_x$), the 2-dim density contains also the $f$-type gluon Sivers function:
\begin{equation}
 \label{eq:rho_Sivers}
 x \rho_{p^\rightarrow} (x, p_x, p_y) = x \, f_1 (x, \pT^2)
 + x \frac{p_y}{M} \, f_{1T}^{\perp} (x, \pT^2)
 \; . 
\end{equation}
The lower panels of Fig.~\ref{fig:rho_unpol_Sivers} show such density in the same conditions and with the same notation as before. Since the nucleon is polarized along the $\hat{x}$ axis, the contour plot shows a distortion along the $\hat{y}$ axis. The asymmetry is clearly visible at $x=10^{-1}$ (right panel), and it is emphasized by the ancillary 1-dim plot at $p_x = 0$. The distortion fades away for decreasing $x$, as shown in the left panel at $x=10^{-3}$. 

If we consider the gluon also in a linearly polarized state, then the 2-dim densities for various nucleon polarizations can become more complicated. The simplest case is for a unpolarized nucleon: the $x \rho^{\leftrightarrow}$ is a linear combination of the T-even gluon TMDs $f_1$ and $h_1^\perp$, and it has been studied in Ref.~\cite{Bacchetta:2020vty} (see lower panels of Fig.5 there). If the nucleon has a longitudinal polarization $S_L$, the 2-dim density $x \rho^{\leftrightarrow}_{p^\odot}$ is a linear combination of $f_1$, $h_1^\perp$ and $h_{1L}^{\perp}$. Finally, if the nucleon has transverse polarization $S_T$ the 2-dim density $x \rho^{\leftrightarrow}_{p^\rightarrow}$ is a linear combination of $f_1$, $h_1^\perp$, $f_{1T}^{\perp}$, $h_1$ and $h_{1T}^{\perp}$, the latter two ones entering with different coefficients depending on the relative angle between the nucleon and gluon polarizations. Apart for the case of unpolarized nucleon $x \rho^\leftrightarrow$, the other 2-dim densities are thus superpositions of three or more gluon TMDs, and their probabilistic interpretation becomes more involved. 

Therefore, we prefer to isolate each T-odd TMD for linearly polarized gluons using the projectors discussed in Sec.~\ref{ss:projectors}, and we plot them for nucleon polarizations along specific directions. 

We first select the nucleon longitudinally polarized along its direction of motion towards the reader, and the gluon linearly polarized along $\hat{x}$. Using the gluon-gluon correlator $\Phi^{x x}(S_L)$ in Eq.(54) of Ref.~\cite{Meissner:2007rx}, the combination $\Phi^{x x}(S_L) - \Phi^{x x}(-S_L)$ isolates the term $h_{1L}^\perp \, p_x p_y / 2 M^2$. In the upper panels of Fig.~\ref{fig:rho_propeller_butterfly}, we show the contour plot for the $\pT$-distribution of the $f$-type combination $h_{1L}^{\perp}\, p_x p_y / 2 M^2$ from replica 11 at $x=10^{-3}$ (left) and $x=10^{-1}$ (right) and at the scale $Q_0 = 1.64$~GeV, scaled by a factor $10^2$. Because of the $p_x p_y$ weight, the contour plot shows symmetric oscillations along the $p_y = \pm p_x$ directions, emphasized in the 1-dim ancillary plots and becoming more sizeable at $x = 10^{-3}$. Sometimes in the literature, the function $h_{1L}^{\perp}$ is called ``T-odd worm-gear" or ``pseudo worm-gear" in analogy with the corresponding quark function. However, we think that this nomenclature does not capture the main characteristics of this function as it emerges from the upper panels of Fig.~\ref{fig:rho_propeller_butterfly}. Since the nucleon is spinning around a direction pointing towards the reader, because of the displayed quadrupolar shape we propose for $h_{1L}^\perp$ the name of ``propeller" function. 

If we keep the gluon linearly polarized along $\hat{x}$ but we consider the combination $\Phi^{x x}(S_x) - \Phi^{x x}(-S_x)$, we can isolate the term $h_{1T}^\perp \, p_x^2 p_y / 2 M^3$. In the central panels of Fig.~\ref{fig:rho_propeller_butterfly}, we show the contour plot for the $\pT$-distribution of the $f$-type combination $h_{1T}^{\perp} \, p_x^2 p_y / 2 M^3$, scaled by the factor $10^2$ and with the same notations as before. The $p_x^2 p_y$ weight produces oscillations symmetric with respect to the $\hat{y}$ axis, emphasized in the 1-dim ancillary plots with slightly displaced slices at $p_y=-0.1$ GeV and $p_x = 0.1$ GeV. The T-odd gluon TMD $h_{1T}^\perp$ is sometimes referred to as ``pretzelosity" in analogy with the quark case. As for $h_{1L}^\perp$, we think that this nomenclature is misleading. The peculiar shape of the contour plot in the lower panels of Fig.~\ref{fig:rho_propeller_butterfly} suggests for $h_{1T}^{\perp}$ the name of ``butterfly" function. 

Finally, if we turn the gluon linear polarization along the $\hat{y}$ axis but keeping the nucleon polarization along $\hat{x}$, the combination $\Phi^{y y}(S_x) - \Phi^{y y}(-S_x)$ isolates the linearity function through the term $- h_1 \, p_y/2M$. In the lower panels of Fig.~\ref{fig:rho_propeller_butterfly}, we show the contour plot for the $\pT$-distribution of the $f$-type combination $- h_1 \, p_y/2M$, scaled by the factor $10$ and with the same notations as before. The $p_y$ weight produces oscillations symmetric with respect to the $\hat{x}$ axis, emphasized in the 1-dim ancillary plots with slightly displaced slices at $p_y=-0.1$ GeV and $p_x = 0.1$ GeV.




\section{Summary and Outlook}
\label{s:conclusions}


In this paper, we have presented a model calculation of all four leading-twist T-odd gluon TMDs within a spectator model approach, providing 
insights into the complex interplay among gluon transverse momentum, gluon polarization, and nucleon spin, and offering a detailed (model-dependent) picture of the distribution of gluons in the nucleon.
This paper completes our previous work~\cite{Bacchetta:2020vty}, where we computed all the leading-twist T-even gluon TMDs in the same framework. 

The model is based on the idea that a nucleon can split into a gluon and remainders that are treated as a single spectator fermion. This spectator mass is allowed to vary within a continuous range, described by a spectral function. Non vanishing T-odd structures are generated by the interference between the tree-level amplitude and an amplitude with final-state interactions, which in our model are approximated as a single-gluon exchange between the gluon and the spectator. The structure of interaction vertices reflects the nature of the involved particles. Since the spectator has spin-$\textstyle{\frac{1}{2}}$, the vertices are modeled resembling the free nucleon electromagnetic current, replacing the Dirac and Pauli form factors with dipolar functions $g_1(p^2)$ and $g_2(p^2)$. For sake of simplicity, all model parameters have been kept the same as in our previous work on T-even gluon TMDs, where they were fixed by fitting the transverse-momentum-integrated gluon TMDs onto known parametrizations of the corresponding collinear unpolarized and helicity gluon PDFs at the lowest scale $Q_0 = 1.64$ GeV~\cite{Bacchetta:2020vty}. 


As it is well known, gluon TMDs have a more intricate dependence on the structure of the color flow (gauge link), which in turn introduces a dependence on the involved process. There are two main classes of gluon TMDs, the so-called Weizs\"acker--Williams (WW) gluon TMDs (also called $f$-type) and the dipole gluon TMDs (also called $d$-type). In general, the two classes cannot be connected, as the WW and dipole gluon TMDs carry different physical information and appear in different processes. Due to the simplifying assumptions in our model, the differences between $f$-type and $d$-type gluon TMDs amount only to a calculable color factor: the size of the $d$-type gluon TMDs is 5/9 of the $f$-type ones.

We have provided analytical and numerical results for the $f$-type T-odd gluon TMDs using two versions of the model: a simpler version with a single form factor ($g_1$, taking $g_2 = 0$) for the nucleon-gluon-spectator and spectator-gluon-spectator vertices, and the full calculation with both $g_1$ and $g_2$ form factors. In the first case, we obtain nonvanishing results only for the  Sivers ($f_{1T}^{\perp}$) and linearity ($h_1$) functions. They turn out to be much smaller than the T-even unpolarized TMD ($f_1$), and they show a decreasing trend for smaller values of $x$. In the full calculation, this trend is reversed and the size becomes comparable to $f_1$, suggesting that sizeable asymmetries generated by such functions could be measurable at small $x$. Moreover, we obtain non vanishing results also for the other two T-odd gluon TMDs: the $h_{1L}^\perp$ (which we name ``propeller") and the $h_{1T}^\perp$ (which we name ``butterfly"). However, both functions have a very small size, particularly the butterfly function, casting some doubts on the actual possibility of ever extracting them from measured spin asymmetries. We computed the T-odd $f$-type gluon TMDs also in the quark-target model. Only the Sivers and linearity functions are different from zero, and their expression matches known results in the literature.

As a final remark, the magnitude of the T-odd gluon TMDs crucially depends on the model parameters. For sake of simplicity, in this paper we have taken them equal to the model parameters of the T-even gluon TMDs~\cite{Bacchetta:2020vty}. However, our model is flexible enough to account for different couplings and different color structures in the interaction vertices, such that the differences between $f$-type and $d$-type gluon TMDs would not amount to a simple color factor. Only future data from the Electron-Ion Collider~\cite{Boer:2011fh,Accardi:2012qut,AbdulKhalek:2021gbh,Khalek:2022bzd,Abir:2023fpo} and new-generation machines~\cite{Arbuzov:2020cqg,Chapon:2020heu,Amoroso:2022eow,Anchordoqui:2021ghd,Feng:2022inv,Accardi:2023chb} will help us to overcome this limitations, and explore also the intriguing connections between our 
polarized gluon TMDs at small-$x$ and the small-$x$ unintegrated gluon density within a hybrid high-energy and collinear factorization framework (see, e.g., Refs.~\cite{Celiberto:2022rfj,Deak:2018obv,Silvetti:2022hyc}).

\begin{acknowledgments}
This work is supported by the European Research Council (ERC) under the European Union's Horizon 2020 research and innovation program (grant agreement
No. 647981, 3DSPIN), by the Italian MIUR under the FARE program (code n. R16XKPHL3N, 3DGLUE), by the Atracci\'on de Talento Grant (n. 2022-T1/TIC-24176), 
and by the European Union ``Next Generation EU'' program through the Italian PRIN 2022 grant
n.~20225ZHA7W.
F.G.C. thanks the Universit\`a degli Studi di Pavia for the warm hospitality.
\end{acknowledgments}

\appendix

\section{Master integrals}
\label{a:integrals}

Here below, we list the master integrals involved in the expressions of our T-odd $f$-type gluon TMDs. We first define 
\begin{equation}
 \label{a:tanhm1}
 T_h(|\pT|) = \tanh^{-1} \sqrt{ \frac{\pT^2}{\pT^2 + 4 \LXtwoL} } \; .
\end{equation}
Then, we have
\begin{equation}  
\begin{split}
 {\cal D}_1 (p) &= \frac{1}{2P^+} \int \frac{d^2 \lT}{(2\pi)^2} \frac{1}{[\lT^2 + \LXtwoL]^2} \, \frac{1}{[(\lT + \pT)^2 + \LXtwoL]^2}
 \\ 
 &=  \frac{1}{8 \pi P^+} \left[ 2 \frac{1 - 2 \LXtwoL/\pT^2}{\LXtwoL \, [\pT^2 + 4 \LXtwoL]^2} + 16 \frac{\pT^2 + \LXtwoL}{|\pT|^3 \, [\pT^2 + 4 \LXtwoL]^{\nicefrac{5}{2}}} \, T_h(|\pT|) \right] \;,
\label{a:D1}
\end{split} \end{equation}

\begin{equation}  \begin{split}
 {\cal D}_2 (p) &= \frac{1}{2P^+} \int \frac{d^2 \lT}{(2\pi)^2} \frac{\lT \cdot \pT}{\pT^2}\frac{1}{[\lT^2 + \LXtwoL]^2} \, \frac{1}{[(\lT + \pT)^2 + \LXtwoL]^2}
 \equiv - \frac{1}{2} \, {\cal D}_1 (p)
\label{a:D2}
\end{split} \end{equation}

\begin{equation}  \begin{split}
 {\cal D}_3 (p) &= \frac{1}{2P^+} \int \frac{d^2 \lT}{(2\pi)^2} \frac{(\lT \cdot \pT)^2}{|\pT|^4}\frac{1}{[\lT^2 + \LXtwoL]^2} \, \frac{1}{[(\lT + \pT)^2 + \LXtwoL]^2}
 \\ 
 &= \frac{1}{8 \pi P^+} \left[\frac{1 + 4 \LXtwoL/\pT^2 + 12 \LXfourL/\pT^4}{\LXtwoL \, [\pT^2 + 4 \LXtwoL]^2} - 24 \frac{[\pT^2 + 2 \LXtwoL] \, \LXtwoL}{|\pT|^5 \, [\pT^2 + 4 \LXtwoL]^{\nicefrac{5}{2}}} \, T_h(|\pT|) \right] \;,
\label{a:D3}
\end{split} \end{equation}

\begin{equation}  \begin{split}
 {\cal D}_4 (p) &= \frac{1}{2P^+} \int \frac{d^2 \lT}{(2\pi)^2} \frac{\lT^2}{\pT^2}\frac{1}{[\lT^2 + \LXtwoL]^2} \, \frac{1}{[(\lT + \pT)^2 + \LXtwoL]^2}
 \\ 
 &= \frac{1}{8 \pi P^+} \left[ \frac{1 + 2 \LXtwoL/\pT^2 + 4 \LXfourL/\pT^4}{\LXtwoL \, [\pT^2 + 4 \LXtwoL]^2} + 4 \frac{\pT^4 - 4 \LXfourL}{|\pT|^5 \, [\pT^2 + 4 \LXtwoL]^{\nicefrac{5}{2}}} \, T_h(|\pT|) \right] \;,
\label{a:D4}
\end{split} \end{equation}

\begin{equation}  \begin{split}
 {\cal D}_5 (p) &= \frac{1}{2P^+} \int \frac{d^2 \lT}{(2\pi)^2} \frac{\lT \cdot \pT}{\pT^2} \frac{\lT^2}{\pT^2}\frac{1}{[\lT^2 + \LXtwoL]^2} \, \frac{1}{[(\lT + \pT)^2 + \LXtwoL]^2}
 \\ 
 &= - \frac{1}{8 \pi P^+} \left[ \frac{1 + 6 \LXtwoL/\pT^2 + 14 \LXfourL/\pT^4}{\LXtwoL \, [\pT^2 + 4 \LXtwoL]^2} - 2 \frac{\pT^4 + 14 \LXtwoL \, \pT^2 + 28 \LXfourL}{|\pT|^5 \, [\pT^2 + 4 \LXtwoL]^{\nicefrac{5}{2}}} \, T_h(|\pT|) \right] \;,
\label{a:D5}
\end{split} \end{equation}

\begin{equation}  \begin{split}
 {\cal D}_6 (p) &= \frac{1}{2P^+} \int \frac{d^2 \lT}{(2\pi)^2} \frac{|\lT|^4}{|\pT|^4}\frac{1}{[\lT^2 + \LXtwoL]^2} \, \frac{1}{[(\lT + \pT)^2 + \LXtwoL]^2}
 \\ 
 &= \frac{1}{8 \pi P^+} \bigg[ \frac{[\pT^2 + 2 \LXtwoL] \, [\pT^4 + 8 \LXtwoL \, \pT^2 + 22 \LXfourL]}{\LXtwoL \, \pT^6 \, [\pT^2 + 4 \LXtwoL]^2}
 \\ 
 &\quad - 4 \frac{\pT^6 + 12 \LXtwoL \, \pT^4 + 40 \LXfourL \, \pT^2 + 44 \LXsixL}{|\pT|^7 \, [\pT^2 + 4 \LXtwoL]^{\nicefrac{5}{2}}} \, T_h(|\pT|) \bigg] \;,
\label{a:D6}
\end{split} \end{equation}

\begin{equation}  \begin{split}
 {\cal D}_7 (p) &= \frac{1}{2P^+} \int \frac{d^2 \lT}{(2\pi)^2} \frac{(\lT \cdot \pT)^2}{|\pT|^4} \frac{|\lT|^2}{\pT^2}\frac{1}{[\lT^2 + \LXtwoL]^2} \, \frac{1}{[(\lT + \pT)^2 + \LXtwoL]^2}
 \\ 
 &= \frac{1}{8 \pi P^+} \bigg[ \frac{|\pT|^7 + 10 \LXtwoL \, |\pT|^5 + 36 \LXfourL \, |\pT|^3 + 36 \LXsixL \, |\pT|}{\LXtwoL \, |\pT|^7 \, [\pT^2 + 4 \LXtwoL]^2}
 \\ 
 &\quad - 6 \frac{[\pT^2 + 2 \LXtwoL]^2 \, [\pT^2 + 6 \LXtwoL]}{|\pT|^7 \, [\pT^2 + 4 \LXtwoL]^{\nicefrac{5}{2}}} \, T_h(|\pT|) \bigg] \;,
\label{a:D7}
\end{split} \end{equation}

\begin{equation}  \begin{split}
 {\cal D}_8 (p) &= \frac{1}{2P^+} \int \frac{d^2 \lT}{(2\pi)^2} \frac{(\lT \cdot \pT)^3}{|\pT|^6} \frac{1}{[\lT^2 + \LXtwoL]^2} \, \frac{1}{[(\lT + \pT)^2 + \LXtwoL]^2}
 \\ 
 &= - \frac{1}{8 \pi P^+} \bigg[ \frac{1 + 7 \LXtwoL/\pT^2 + 18 \LXfourL/\pT^4}{\LXtwoL \, [\pT^2 + 4 \LXtwoL]^2}
 - 4 \frac{\pT^4 + 10 \LXtwoL \, \pT^2 + 18 \LXfourL}{|\pT|^5 \, [\pT^2 + 4 \LXtwoL]^{\nicefrac{5}{2}}} \, T_h(|\pT|) \bigg] \;.
\label{a:D8}
\end{split} \end{equation}


\section{Full calculation of gluon TMDs: The  coefficients $C_{ijk}^{[F], l}$}
\label{a:coeffs}

In the following, we list the final expressions of the $C_{ijk}^{[F], l}$ coefficients in Eqs.~\eqref{eq:F_lin_comb} and \eqref{eq:C_lin_comb} for each T-odd gluon TMD $F$ and for $l=1,..,8, \  i,j,k=1,2$. 
We note that 
the $C_{111}^{[F],2}$ coefficients for the $f$-type gluon Sivers ($[F] \equiv \left[f_{1T}^\perp\right]$) and linearity ($[F] \equiv \left[h_1\right]$) 
have already been derived when discussing the computation of these T-odd gluon TMDs 
in the $g_1$-vertex approximation (see Eqs.~\eqref{f1Tp_pp_g1_2} and~\eqref{h1_pp_g1_2}, respectively).

\subsection{Sivers function $f_{1T}^{\perp}$}
\label{ss:Sivers_f_full}

In Tabs.~\ref{tab:SIV1}-\ref{tab:SIV4}, we list the coefficients $C_{ijk}^{\left[f_{1T}^\perp\right],l}$ for $i,j,k=1,2$ and $l=1,..,8$. 

\begingroup
\setlength{\tabcolsep}{12pt} 
\renewcommand{\arraystretch}{3.0}

\begin{table}[H]
\centering
\caption{Coefficients functions of the $f$-type gluon Sivers TMD for $ijk=\{111,112\}$.}
\label{tab:SIV1}
\scriptsize
\begin{tabular}[c]{|c|c|c|}
\hline

$C_{ijk}^{\left[f_{1T}^\perp\right],l}$ & $ijk=111$ & $ijk=112$ \\
\hline

$l=1$ & $0$ & \(\displaystyle\frac{12}{x}\)$\left[ 2x \left( M^2 (2 - x) x - (M - M_X)^2 \right) + 2 \pT^2 (2-x) \right]$ \\
\hline

$l=2$ & $-48 M (1-x) \left[ M_X - M (1-x) \right] $ & \(\displaystyle - \frac{12}{x}\)$\left[ x \left( M (1-x) + M_X \right) \, \left( M (1-x)^2 + M_X (1 + x) \right) - \pT^2 (x^2-5x+8) \right]$ \\
\hline

$l=3$ & $0$ & \(\displaystyle\frac{24}{x}\)$\, \pT^2 \, (x^2-2x+2)$ \\
\hline

$l=4$ & $0$ & $12 \, \pT^2 \, (1-x)$ \\
\hline

$l=5$ & $0$ & $0$ \\
\hline

$l=6$ & $0$ & $0$ \\
\hline

$l=7$ & $0$ & $0$ \\
\hline

$l=8$ & $0$ & $0$ \\

\hline
\end{tabular}
\end{table}

\begin{table}[H]
\centering
\caption{Coefficients functions of the $f$-type gluon Sivers TMD for $ijk=\{121,122\}$.}
\label{tab:SIV2}
\scriptsize
\begin{tabular}{|c|c|c|}
\hline

$C_{ijk}^{\left[f_{1T}^\perp\right],l}$ & $ijk=121$ & $ijk=122$ \\
\hline

$l=1$ & $0$ & $0$ \\
\hline

$l=2$ & \(\displaystyle\frac{6}{x}\)$\, \left[ - M^2 (2-x) \, x \, (1-x)^2 + M_X^2 \, (2-x) \, x + \pT^2 \, (x^2-10x+8) \right]$ & $3x\, $\(\displaystyle\frac{M-M_X}{M}\)$\, \left[ M^2 (1-x)^2 - M_X^2 + \pT^2\right ]$ \\
\hline

$l=3$ & \(\displaystyle\frac{48 \pT^2}{x}\)$\, (1-x)$ & $0$ \\
\hline

$l=4$ & $-6 \pT^2 \, (2-x)$ & \(\displaystyle\frac{3\pT^2}{M}\)$\, \left[ M \, (4-3x) - M_X \, (x+4)\right]$ \\
\hline

$l=5$ & $0$ & \(\displaystyle\frac{12 \pT^2}{M}\)$\, \left[ M (1-x) - M_X \right]$ \\
\hline

$l=6$ & $0$ & $0$ \\
\hline

$l=7$ & $0$ & $0$ \\
\hline

$l=8$ & $0$ & $0$ \\

\hline
\end{tabular}
\end{table}

\begin{table}[H]
\centering
\caption{Coefficients functions of the $f$-type gluon Sivers TMD for $ijk=\{211,212\}$. 
}
\label{tab:SIV3}
\scriptsize
\begin{tabular}{|c|c|c|}
\hline

$C_{ijk}^{\left[f_{1T}^\perp\right],l}$ & $ijk=211$ & $ijk=212$ \\
\hline

$l=1$ & \(\displaystyle - \frac{24}{x}\)$\, \left[ x\, \left( M_X^2 - M^2(1-x)^2 \right) + (2-x)\, \pT^2\right]$ & $0$ \\
\hline

$l=2$ & $\tarr c \displaystyle - \frac{6}{(1-x)x} \, \left[ x^2 \, \left( M_X-M(1-x) \right) \, \left( M\, (1-x)\, x + M_X \, (4-3x) \right) \right. \\ \left. - \pT^2 \, (x^3-8x+8) \right] \earr$ & $\tarr c \displaystyle - \frac{3}{M(1-x)} \, \left[ M^3 \, (2-x) \, (1-x)^2 \right. \\ \left. + 3M^2\, M_X\, (1-x)^2 \, x^2 \right. \\ \left. - M \, \left( M_X^2 \, x \, (3x^2-4x+2) + \pT^2\, (x^2-6x+4) \right) \right.\\ \left. + M_X^3\, x^2 + M_X\, \pT^2 \, (3x^2-8x+4) \right] \earr$ \\
\hline

$l=3$ & $0$ & \(\displaystyle \frac{12 \pT^2}{M} \)$ \, \left[ M_X \, (1-x) - M \right]$ \\
\hline

$l=4$ & $6 \pT^2 \, $ \(\displaystyle \frac{(2-x)^2}{(1-x)}\) & \(\displaystyle \frac{3 x\pT^2}{M(1-x)}\)$\, \left[ M \, (2-x) - M_X \, x \right]$ \\
\hline

$l=5$ & $0$ & $0$ \\
\hline

$l=6$ & $0$ & $0$ \\
\hline

$l=7$ & $0$ & $0$ \\
\hline

$l=8$ & $0$ & $0$ \\

\hline
\end{tabular}
\end{table}

\begin{table}[H]
\centering
\caption{Coefficients functions of the $f$-type gluon Sivers TMD for $ijk=\{221,222\}$. 
}
\label{tab:SIV4}
\scriptsize
\begin{tabular}{|c|c|c|}
\hline

$C_{ijk}^{\left[f_{1T}^\perp\right],l}$ & $ijk=221$ & $ijk=222$ \\
\hline

$l=1$ & $0$ & $0$ \\
\hline

$l=2$ & $\tarr c 3x\,$\(\displaystyle \frac{M^2(1-x)^2 - M_X^2 + \pT^2}{2M(1-x)} \)$ \\ \times \left[ M \, (1-x)\, (7-2x) - M_X\, (7-3x) \right] \earr$ & $\tarr c \displaystyle \frac{3}{8M^2(1-x)x} \, \left[ x^2 \, \left( M^2\, (1-x)^2-M_X^2 \right) \right.\\ \left. \times \left( M^2\, (2 x^3+x^2-3)+4M\, M_X\, (3-x) \, x+M_X^2\, (3-2 x)\right) \right.\\ \left. - 4x\pT^2 \, \left( M^2 \, (1-x) \, (x^3 - 5 x + 6) - M \, M_X \, (3-x) x^2 \right. \right.\\ \left. \left. + M_X^2 \, (x^2+7x-6) \right) \right. \\ \left. + \pT^4 \, (2x^3-3x^2-24x+32) \right] \earr$ \\
\hline

$l=3$ & \(\displaystyle \frac{12\pT^2}{M} \)$\, \left[ M_X - M\, (1-x)\right]$ & $\tarr c \displaystyle - \frac{3\pT^2}{4M^2\,(1-x)\,x} \,\left[ M^2\, (1-x) \, x\, (2x^3+5x^2-13x+4) \right.\\ \left. + 8M\, M_X\, (1-x) \, x^2 + M_X^2\, x\, (2x^2+11x-4) \right.\\ \left. - \pT^2\, (2x^3+3x^2-28x+32) \right] \earr$ \\
\hline

$l=4$ & \(\displaystyle \frac{3x\,\pT^2}{2M\,(1-x)} \)$\, \left[ M\, (1-x)\, (7-2x) - M_X \, (7-3x)\right]$ & $\tarr c \displaystyle - \frac{3\pT^2}{8M^2(1-x)} \, \left[ M^2\, (1-x)\, (2x^3-5x^2-x+16) \right.\\\left. + 4M\, M_X\, x\, (x^2+x-4) + M_X^2\, x\, (2x^2+13x-6) \right.\\\left. - \pT^2\, (2x^2-x+8)\right] \earr$ \\
\hline

$l=5$ & $0$ & $\tarr c \displaystyle \frac{3\pT^2}{8M^2(1-x)} \, \left[ M^2 \, (1-x)^2 \, (2x^2-x-8) \right.\\ \left. + 16 M \, M_X \, (1-x)\, x + M_X^2 \, (2x^2+15x-8) \right. \\ \left. + \pT^2 \, (6x^2-11x+32) \right] \earr$ \\
\hline

$l=6$ & $0$ & \(\displaystyle \frac{3\pT^4}{8M^2} \)\(\displaystyle \frac{2x^2-x+8}{1-x} \) \\
\hline

$l=7$ & $0$ & $0$ \\
\hline

$l=8$ & $0$ & \(\displaystyle \frac{6\pT^4}{M^2} \) \(\displaystyle \frac{2-x}{x} \) \\

\hline
\end{tabular}
\end{table}

\endgroup

\subsection{Linearity function $h_1$}
\label{ss:linearity_f_full}

In Tabs.~\ref{tab:LIN1}-\ref{tab:LIN4}, we list the coefficients $C_{ijk}^{\left[h_1\right],l}$ for $i,j,k=1,2$ and $l=1,..,8$. 

\begingroup
\setlength{\tabcolsep}{12pt} 
\renewcommand{\arraystretch}{3.0}

\begin{table}[H]
\centering
\caption{Coefficients functions of the $f$-type gluon linearity for $ijk=\{111,112\}$.}
\label{tab:LIN1}
\scriptsize
\begin{tabular}{|c|c|c|}
\hline

$C_{ijk}^{\left[h_1\right],l}$ & $ijk=111$ & $ijk=112$ \\
\hline

$l=1$ & $0$ & $\tarr c \displaystyle \frac{6}{x} \, \left[ x\, (4-3x)\, \left( M^2\, (1-x)^2-M_X^2\right) + \pT^2\, (3x^2-12x+8) \right] \earr$ \\
\hline

$l=2$ & $- 96 M \, \left[ M_X - M\, (1-x)\right]$ & $\tarr c \displaystyle -\frac{6}{x} \, \left[ x\, \left( M^2\, (3x^3-6x^2+11x-8) + 8M\, M_X\, x - M_X^2\, (8-3x)\right) - \pT^2\, (3x^2-8x+16)\right] \earr$ \\
\hline

$l=3$ & $0$ & $48\pT^2$ \(\displaystyle \frac{1-x}{x} \) \\
\hline

$l=4$ & $0$ & $48\pT^2$ \\
\hline

$l=5$ & $0$ & $0$ \\
\hline

$l=6$ & $0$ & $0$ \\
\hline

$l=7$ & $0$ & $0$ \\
\hline

$l=8$ & $0$ & $0$ \\

\hline
\end{tabular}
\end{table}

\begin{table}[H]
\centering
\caption{Coefficients functions of the $f$-type gluon linearity for $ijk=\{121,122\}$.}
\label{tab:LIN2}
\scriptsize
\begin{tabular}{|c|c|c|}
\hline

$C_{ijk}^{\left[h_1\right],l}$ & $ijk=121$ & $ijk=122$ \\
\hline

$l=1$ & $0$ & $0$ \\
\hline

$l=2$ & $\tarr c \displaystyle \frac{6}{x} \, \left[ x\, \left( M^2\, (1-x)^2\, (2+3x) - 16M\, M_X\, (1-x)\,x \right. \right.\\\left. \left. - M_X^2\, (2-13x)\right) + \pT^2\, (5x^2-2x+8)\right] \earr$ & $\tarr c \displaystyle \frac{3}{M} \, \left[ M^2\, M_X\, x\, (7x^2+2x-9) + M\, x\, \left( 5M^2\, (1-x)^2+11\pT^2\right) \right.\\\left. - M\,M_X^2\,x\,(5-16x) + 9M_X^3\,x + M_X\,\pT^2\,(8+x)\right] \earr$ \\
\hline

$l=3$ & $\tarr c \displaystyle \frac{48\pT^2}{x} \earr$ & $\tarr c \displaystyle \frac{6\pT^2}{M} \, \left[ 4M\, x + M_X\, (10+x)\right] \earr$ \\
\hline

$l=4$ & $0$ & $\tarr c \displaystyle \frac{6\pT^2}{M} [4M(1-x) + M_X(2+x)] \earr$ \\
\hline

$l=5$ & $0$ & $\tarr c \displaystyle \frac{6\pT^2}{M} \, \left[ 4M\, (1-x) + M_X\, (2-3x)\right] \earr$ \\
\hline

$l=6$ & $0$ & $0$ \\
\hline

$l=7$ & $0$ & $0$ \\
\hline

$l=8$ & $0$ & $\tarr c \displaystyle - \frac{18 \pT^2 M_X}{M} \, (2-x) \earr$  \\

\hline
\end{tabular}
\end{table}

\begin{table}[H]
\centering
\caption{Coefficients functions of the $f$-type gluon linearity for $ijk=\{211,212\}$.}
\label{tab:LIN3}
\scriptsize
\begin{tabular}{|c|c|c|}
\hline

$C_{ijk}^{\left[h_1\right],l}$ & $ijk=211$ & $ijk=212$ \\
\hline

$l=1$ & $\tarr c \displaystyle -\frac{6}{x} \, \left[ \left( M^2\, (1-x)^2-M_X^2\right)\, (4-3x)\, x \right.\\\left. + \pT^2\, (3x^3-12x+8)\right] \earr$ & $-9x\, \left[ M^2\, (1-x)^2 - M_X^2 - \pT^2\right]$ \\
\hline

$l=2$ & $\tarr c \displaystyle -\frac{6}{(1-x)x} \, \left[ x^2\, \left( M^2\, (1-x)^2\, (7-2x) \right. \right.\\\left. \left. - 8M\, M_X\,(1-x) + M_X^2\, (1+2x)\right) \right. \\\left. + \pT^2\, (2x^3-7x^2+16x-8)\right] \earr$ & $\tarr c \displaystyle -\frac{3}{M(1-x)} \, \left[ M^3\, (1-x)^2\, x\, (9-4x) \right. \\ \left. + M^2\, M_X\, (x^3+5x-6) + 3M\, M_X^2\, (4-3x)\, x \right. \\ \left. + M\, \pT^2\, (4x^2-9x+8) + M_X\, (M_X^2+\pT^2)\, (6-x)\right] \earr$ \\
\hline

$l=3$ & $0$ & $-24\pT^2$ \\
\hline

$l=4$ & $0$ & $0$ \\
\hline

$l=5$ & $0$ & $0$ \\
\hline

$l=6$ & $0$ & $0$ \\
\hline

$l=7$ & $0$ & $0$ \\
\hline

$l=8$ & $0$ & $0$ \\

\hline
\end{tabular}
\end{table}

\begin{table}[H]
\centering
\caption{Coefficients functions of the $f$-type gluon linearity for $ijk=\{221,222\}$.}
\label{tab:LIN4}
\scriptsize
\begin{tabular}{|c|c|c|}
\hline

$C_{ijk}^{\left[h_1\right],l}$ & $ijk=221$ & $ijk=222$ \\
\hline

$l=1$ & $0$ & $0$ \\
\hline

$l=2$ & $\tarr c \displaystyle \frac{3}{M(1-x)} \, \left[ 2M_X\, \left( M^2\, x\, (3+2x)\, (1-x)^2 \right. \right. \\ \left. \left. + \pT^2\, (2x^2+5x-4) \right) \right. \\ \left. + M\, (1-x)\, x \, \left( - M^2\, (6-x)\, (1-x)^2 - \pT^2\, (10+x) \right) \right. \\ \left. + M \, M_X^2\, (17x^2-23x+6)\, x - 6M_X^3\, (1-2x)\, x \right] \earr$ & $\tarr c \displaystyle -\frac{3}{4M^2(1-x)x} \, \left[ \pT^2\, x\, \left( M^2\, (1-x)\, (6x^3-31x^2+6x-4) \right. \right. \\ \left. \left. + 2M\, M_X\, (x^2+21x-16)\, x + 2M_X^2\, (2+x)\, (1+3x) \right) \right. \\ \left. - x^2\, \left( M\, (1-x)-M_X \right) \, \left( M^3\, (x^2+x-2)^2 - M_X^3\, (4-17x) \right. \right.\\ \left. \left. + M^2\, M_X\, (3x^2-18x+4)\, (1-x) + M\, M_X\, (17x^2-x-4) \right) \right. \\ \left. + \pT^4\, (5 x^3-30x^2+52x-16) \right] \earr$ \\
\hline

$l=3$ & $-12\pT^2\, (2-x)$ & $\tarr c \displaystyle - \frac{3\pT^2}{2M^2(1-x)x} \, \left[ M^2\, (2x^3-15x^2+8x+2)\, (1-x)\, x \right.\\ \left. - 4M\, M_X\, (1-x)\, x^2 + M_X^2\, (4x^2+5x-2) \right. \\ \left. + \pT^2\, (2x^3-9x^2+30x-16)\right] \earr$ \\
\hline

$l=4$ & $-12x\, \pT^2$ & $\tarr c \displaystyle -\frac{3\pT^2}{4M^2(1-x)} \, \left[ M^2\, (3x^3-7x^2+2x-8)\, (1-x) \right. \\ \left. - 24M\, M_X\, (1-x)\, x - M_X^2\, (x^2+8x-8) \right. \\ \left. + \pT^2\, (3x^2-12x+8)\right] \earr$ \\
\hline

$l=5$ & $0$ & $\tarr c \displaystyle \frac{3\pT^2}{2M^2(1-x)} \, \left[ M^2\, (1-x)^2\, (x^2-7x+2) + 8M\, M_X\, (1-x)\, x \right. \\ \left. - M_X^2\, (7x^2-11x+2) - \pT^2\, (x^2-x-6) \right] \earr$ \\
\hline

$l=6$ & $0$ & $\tarr c \displaystyle \frac{3\pT^4}{M^2} \frac{x}{1-x} \earr$ \\
\hline

$l=7$ & $0$ & $\tarr c \displaystyle -\frac{3\pT^4}{M^2} \frac{x}{1-x} \earr$ \\
\hline

$l=8$ & $0$ & $\tarr c \displaystyle \frac{3\pT^2}{2M^2(1-x)x} \, \left[ 3\, (2-x)\, x\, \left( M_X^2\, (1-2x)-M^2\, (1-x)^2\right) \right. \\ \left. + \pT^2\, (3x^2-18x+8)\right] \earr$ \\

\hline
\end{tabular}
\end{table}

\endgroup


\subsection{Propeller function $h_{1L}^{\perp}$}
\label{ss:TO-WG_f_full}

In Tabs.~\ref{tab:WG1}-\ref{tab:WG4}, we list the coefficients $C_{ijk}^{\left[h_{1L}^\perp \right],l}$ for $i,j,k=1,2$ and $l=1,..,8$. 


\begingroup
\setlength{\tabcolsep}{12pt} 
\renewcommand{\arraystretch}{3.0}

\begin{table}[H]
\centering
\caption{Coefficients functions of the $f$-type gluon propeller for $ijk=\{111,112\}$.}
\label{tab:WG1}
\scriptsize
\begin{tabular}{|c|c|c|}
\hline

$C_{ijk}^{\left[h_{1L}^\perp\right],l}$ & $ijk=111$ & $ijk=112$ \\
\hline

$l=1$ & $0$ & $96(1-x)\, M^2$ \\
\hline

$l=2$ & $0$ & $48M\, [3M\, (1-x)-M_X]$ \\
\hline

$l=3$ & $0$ & $96M\, [M\, (1-x)-M_X]$ \\
\hline

$l=4$ & $0$ & $-48M\, [M\, (1-x)-M_X]$ \\
\hline

$l=5$ & $0$ & $0$ \\
\hline

$l=6$ & $0$ & $0$ \\
\hline

$l=7$ & $0$ & $0$ \\
\hline

$l=8$ & $0$ & $0$ \\

\hline
\end{tabular}
\end{table}

\begin{table}[H]
\centering
\caption{Coefficients functions of the $f$-type gluon propeller for $ijk=\{121,122\}$.}
\label{tab:WG2}
\scriptsize
\begin{tabular}{|c|c|c|}
\hline

$C_{ijk}^{\left[h_{1L}^\perp\right],l}$ & $ijk=121$ & $ijk=122$ \\
\hline

$l=1$ & $0$ & $0$ \\
\hline

$l=2$ & $96(1-x)\, M\, [M(1-x)-M_X]$ & $-24\, \left[M \, (1-x) \, \left( M \, (1-x)-2 M_X \, x \right)-M_X^2\, (1-2x)+\pT^2 \right]$ \\
\hline

$l=3$ & $96(1-x)\, M\, [M\, (1-x)-M_X]$ & \(\displaystyle \frac{48}{x} \)$\, \left[ M \, M_X \, (1-x) x^2 - M_X^2 \, x^2 + \pT^2 \, (1-x) \right]$ \\
\hline

$l=4$ & $-48(1-x)\, M\, [M\, (1-x)-M_X]$ & \(\displaystyle - \frac{24}{x} \)$\, \left[ M \, M_X \, (1-x) x^2 - M_X^2 x^2 + 2 \pT^2\right]$ \\
\hline

$l=5$ & $0$ & $-24\pT^2$ \\
\hline

$l=6$ & $0$ & $0$ \\
\hline

$l=7$ & $0$ & $0$ \\
\hline

$l=8$ & $0$ & $0$ \\

\hline
\end{tabular}
\end{table}

\begin{table}[H]
\centering
\caption{Coefficients functions of the $f$-type gluon propeller for $ijk=\{211,212\}$.}
\label{tab:WG3}
\scriptsize
\begin{tabular}{|c|c|c|}
\hline

$C_{ijk}^{\left[h_{1L}^\perp\right],l}$ & $ijk=211$ & $ijk=212$ \\
\hline

$l=1$ & $-96(1-x)\, M^2$ & $0$ \\
\hline

$l=2$ & $-48M\, [M\, (1-x) + M_X]$ & $-48\, [M\, (1-x) - M_X]$ \\
\hline

$l=3$ & $0$ & \(\displaystyle -\frac{48}{x} \)$\, \left[ x \, (M+M_X)\, \left( M\, (1-x)-M_X \right) - \pT^2 \right]$ \\
\hline

$l=4$ & $0$ & \(\displaystyle\frac{24}{x} \)$\, \left[ x \, (M+M_X)\, \left( M\, (1-x)-M_X\right) - 2\pT^2\right]$ \\
\hline

$l=5$ & $0$ & $0$ \\
\hline

$l=6$ & $0$ & $0$ \\
\hline

$l=7$ & $0$ & $0$ \\
\hline

$l=8$ & $0$ & $0$ \\

\hline
\end{tabular}
\end{table}

\begin{table}[H]
\centering
\caption{Coefficients functions of the $f$-type gluon propeller for $ijk=\{221,222\}$.}
\label{tab:WG4}
\scriptsize
\begin{tabular}{|c|c|c|}
\hline

$C_{ijk}^{\left[h_{1L}^\perp\right],l}$ & $ijk=221$ & $ijk=222$ \\
\hline

$l=1$ & $0$ & $0$ \\
\hline

$l=2$ & $\tarr c -24\, \left[ M^2\, (1-x)^2 - 2M\, M_X\, (1-x)\, x \right. \\ \left. - M_X^2\, (1-2x) + \pT^2 \right] \earr$ & \(\displaystyle \frac{24}{M}\, (M_X+M)\, \left[ x\, M_X\, \left( M_X-M\, (1-x)\right)+\pT^2 \right] \) \\
\hline

$l=3$ & $\tarr c \displaystyle  \frac{48(1-x)}{x} \, \left[ x\, (M+M_X)\, \left( M_X-M\, (1-x)\right) + \pT^2\right] \earr$ & \(\displaystyle \frac{24M_X}{M}\, \left[ x\, (M_X+M)\, \left( M_X-M\, (1-x)\right)+2\pT^2\right] \) \\
\hline

$l=4$ & $\tarr c \displaystyle - \frac{24}{x} \, \left[ (1-x)\, x\, (M+M_X)\, \left( M_X-M(1-x)\right) \right. \\ \left. + \pT^2\, (2-x) \right] \earr$ & $\tarr c \displaystyle - \frac{12}{M}\, \left[ x^2\, M_X\, M\, (M_X+M) + x\, M_X\, (M_X^2-M^2) \right. \\ \left. +(M_X-3M)\, \pT^2 \right] \earr$ \\
\hline

$l=5$ & $0$ & \(\displaystyle \frac{12\pT^2}{M}(M + M_X) \) \\
\hline

$l=6$ & $0$ & $0$ \\
\hline

$l=7$ & $0$ & $0$ \\
\hline

$l=8$ & $0$ & $0$ \\

\hline
\end{tabular}
\end{table}

\endgroup

\subsection{Butterfly function $h_{1T}^{\perp}$}
\label{ss:pretzelosity_f_full}



In Tabs.~\ref{tab:PRETZ1}-\ref{tab:PRETZ4}, we list the coefficients $C_{ijk}^{\left[h_{1T}^\perp \right],l}$ for $i,j,k=1,2$ and $l=1,..,8$. 


\begingroup
\setlength{\tabcolsep}{12pt} 
\renewcommand{\arraystretch}{3.0}

\begin{table}[H]
\centering
\caption{Coefficients functions of the $f$-type gluon butterfly for $ijk=\{111,112\}$.}
\label{tab:PRETZ1}
\scriptsize
\begin{tabular}{|c|c|c|}
\hline

$C_{ijk}^{\left[h_{1T}^\perp\right],l}$ & $ijk=111$ & $ijk=112$ \\
\hline

$l=1$ & $0$ & \(\displaystyle - \frac{M^2}{2\pT^2(1-x)x}\, \left[ (1-x)^2\, \left( M^2\, x\, (x^2-5x+4) - 2M_X^2\, x + 2\pT^2\, (4-x) \right) + x\, (x^2+x-2)\, (M_X^2+\pT^2) \right] \) \\
\hline

$l=2$ & $0$ & \(\displaystyle - \frac{M^2}{2\pT^2(1-x)x} \, \left[ (1-x)^2 \, \left( M^2\, x\, (x^2-5x+4) - 2M_X^2\, x + 2\pT^2\, (8-x) \right) + x\, (x^2+x-2)\, (M_X^2+\pT^2) \right] \) \\
\hline

$l=3$ & $0$ & $\tarr c \displaystyle - 4M^2 \frac{1-x}{x} \earr$  \\
\hline

$l=4$ & $0$ & $0$ \\
\hline

$l=5$ & $0$ & $0$ \\
\hline

$l=6$ & $0$ & $0$ \\
\hline

$l=7$ & $0$ & $0$ \\
\hline

$l=8$ & $0$ & $0$ \\

\hline
\end{tabular}
\end{table}

\begin{table}[H]
\centering
\caption{Coefficients functions of the $f$-type gluon butterfly for $ijk=\{121,122\}$.}
\label{tab:PRETZ2}
\scriptsize
\begin{tabular}{|c|c|c|}
\hline

$C_{ijk}^{\left[h_{1T}^\perp\right],l}$ & $ijk=121$ & $ijk=122$ \\
\hline

$l=1$ & $0$ & $0$ \\
\hline

$l=2$ & $\tarr c \displaystyle \frac{M^2}{2x\pT^2} \, \left[ (2-x)\, x\, (M_X^2+\pT^2) \right. \\ \left. - (1-x)^2\, \left( M^2\, (2-x)\, x+8\pT^2 \right) \right] \earr $ & $\tarr c \displaystyle \frac{M}{4\pT^2} \, \left[ M^3 \, x (1-x)^2 + M \left( \pT^2\,  (3 x^2-8x+4)-M_X^2 \, x^2\right) \right. \\ \left. - (1-x) \, \left( M^2 \, M_X \, (1-x) \, x + M \, M_X^2 \, x + M \, \pT^2 \, (4-3x) \right. \right. \\ \left. \left. + 4 M_X \, \pT^2 \right) +M_X^3 \, x - M_X \, \pT^2 \, (4-5x) \right] \earr$ \\
\hline

$l=3$ & \(\displaystyle -\frac{4M^2(1-x)^2}{x} \) & \(\displaystyle -\frac{2-3x}{2} MM_X \) \\
\hline

$l=4$ & $0$ & \(\displaystyle -\frac{2-x}{2} M\, M_X \) \\
\hline

$l=5$ & $0$ & \(\displaystyle -\frac{2-x}{2} M\, M_X \) \\
\hline

$l=6$ & $0$ & $0$ \\
\hline

$l=7$ & $0$ & $0$ \\
\hline

$l=8$ & $0$ & \(\displaystyle \frac{2-x}{2} M\, M_X \) \\

\hline
\end{tabular}
\end{table}

\begin{table}[H]
\centering
\caption{Coefficients functions of the $f$-type gluon butterfly for $ijk=\{211,212\}$.}
\label{tab:PRETZ3}
\scriptsize
\begin{tabular}{|c|c|c|}
\hline

$C_{ijk}^{\left[h_{1T}^\perp\right],l}$ & $ijk=211$ & $ijk=212$ \\
\hline

$l=1$ & $\tarr c \displaystyle \frac{M^2}{2x\pT^2} \, \left[ (1-x) \, \left( 4 M^2 \, (1-x)^2 \, x+4 \pT^2 \, (2-x)\right) \right. \\ \left. +3 M^2 \, x^2 \, (1-x)^2-x \, \left( M_X^2 \, (4-x)+3 \pT^2 \, x\right)\right] \earr$ & \( \displaystyle \frac{xM^2}{4\pT^2} \, \left[ M^2\, (1-x)^2-M_X^2-\pT^2 \right] \) \\
\hline

$l=2$ & $\tarr c \displaystyle -\frac{M^2}{\pT^2(1-x)x} \, \left[ (1-x)^2\, \left( M^2\, x^2 - 4\pT^2\, (2-x) \right) \right. \\ \left. -x\, \left( M_X^2\, x + \pT^2\, (4x^2-7x+4) \right) \right] \earr$ & $\tarr c \displaystyle \frac{M}{4(1-x)\pT^2} \, \left[ (1-x)^2 \, \left( M^3 \, x \, (2x^2-4 x+3) + 2M\, \pT^2\, (2-x)\right) \right. \\ \left. + 2M^3\, x^2\, (1-x)^3 - M_X\, (2-x)\, (1-x)\, \left( 2\pT^2+M^2\, x\, (1-x)\right) \right. \\ \left. - M \, M_X^2\, x\, (3-2x) + M\, \pT^2\, (2 x^3-6x^2+7x-4) \right. \\ \left. + M_X^3\, (2-x)\, x + M_X\, \pT^2\, (x^2+4x-4)\right] \earr$ \\
\hline

$l=3$ & $0$ & $-2MM_X$ \\
\hline

$l=4$ & $0$ & $0$ \\
\hline

$l=5$ & $0$ & $0$ \\
\hline

$l=6$ & $0$ & $0$ \\
\hline

$l=7$ & $0$ & $0$ \\
\hline

$l=8$ & $0$ & $0$ \\

\hline
\end{tabular}
\end{table}

\begin{table}[H]
\centering
\caption{Coefficients functions of the $f$-type gluon butterfly for $ijk=\{221,222\}$.}
\label{tab:PRETZ4}
\scriptsize
\begin{tabular}{|c|c|c|}
\hline

$C_{ijk}^{\left[h_{1T}^\perp\right],l}$ & $ijk=221$ & $ijk=222$ \\
\hline

$l=1$ & $0$ & $0$ \\
\hline

$l=2$ & $\tarr c \displaystyle \frac{M}{4(1-x)\pT^2} \, \left[ M^3\, x\, (3-x)\, (1-x)^3 - (1-x) \, \left( 2 M\,  M_X^2\,  x^2 \right. \right. \\ \left. \left. - 2 M \, \pT^2 \, (x^2-4 x+2)-3 M_X^3 \, x+M_X \, \pT^2 \, (4-7 x)\right) \right. \\ \left. - M \, (1-x)^2 \, \left( M_X \, x \, \left( 3 M_X+M \, (3-x) \right)+\pT^2 \, (4-x)\right) \right. \\ \left. + 2 M_X^3 \, x^2 - 2 M_X\, \pT^2 \, (x^2-4 x+2)\right] \earr$ & $\tarr c \displaystyle - \frac{1}{16(1-x)x\pT^2} \, \left[ \pT^2 \, x \, \left( M^2 \, (1-x) \, (2 x^3-3 x^2+8 x-4) \right. \right.  \\ \left. \left. - 2 M \, M_X \, (3-x) \, x^2+2 M_X^2 \, (9 x^2-10 x+2) \right) \right. \\ \left. + x^2 \, \left( M^2\, ( 1-x)^2-M_X^2 \right) \, \left( - M^2 \, (1-x^3) \right. \right. \\ \left. \left. + 2 M \, M_X \, (3-x) \, x + M_X^2 \, (1-x) \right) \right. \\ \left. + \pT^4 \, (x^3-3 x^2+20 x-16)\right] \earr$ \\
\hline

$l=3$ & $-2M M_X$ & $ \tarr c \displaystyle - \frac{1}{8x} \, \left[ M^2 \, x \, (2x^2-x-2) \right. \\ \left. - M_X^2 \, x \, (1+3x) - 2 \pT^2 \, (8-3x)\right] \earr $ \\
\hline

$l=4$ & $0$ & $\tarr c \displaystyle \frac{1}{16(1-x)} \, \left[ M^2 \, (1-x) \, (x^3-x^2+6x-8) \right. \\ \left. + M_X^2 \, (5x^2-12x+8) + \pT^2 \, (x^2+8x-8) \right] \earr$ \\
\hline

$l=5$ & $0$ & $\tarr c \displaystyle - \frac{1}{16(1-x)} \, \left[ (1-x)\, \left( x \, \left( M^2 \, (1-x)^2 - M_X^2\right) \right. \right. \\ \left. \left. + \pT^2 \, (8-x) \right) + (2-x) \, \left( M^2 \, (2-x) \, (1-x)^2 \right. \right. \\ \left. \left. - M_X^2 \, (2-5x) + \pT^2 \, (2+x) \right) \right]\earr$ \\
\hline

$l=6$ & $0$ & $0$ \\
\hline

$l=7$ & $0$ & $0$ \\
\hline

$l=8$ & $0$ & $\tarr c \displaystyle \frac{2-x}{8(1-x)x} \, \left[ M^2\, (1-x)^2\, x - M_X\, (1-2x)\, x \right. \\ \left. - \pT^2\, (4-5x)\right] \earr $\\

\hline
\end{tabular}
\end{table}

\endgroup


\bibliography{references}


\end{document}